%% file: main_passCheck.tex
\documentclass[journal]{IEEEtran}
\usepackage[latin9]{inputenc}
\usepackage{amsfonts}
\usepackage{amsmath}
\usepackage{amssymb}

\usepackage[unicode=true,
bookmarks=false,
pdfborder={0 0 1},colorlinks=false,hidelinks,breaklinks=false]{hyperref}

\makeatletter
\usepackage{longtable}
\usepackage{multicol}
\usepackage{multirow}
\usepackage{graphicx}
\usepackage{cite}
\usepackage{enumitem}
\usepackage{siunitx}
\usepackage[dvipsnames]{xcolor}
\usepackage{algorithm}
\usepackage{algorithmic}

\newcommand{\sss}{\varphi}
\newcommand{\ssr}{\theta}
\newcommand{\omw}{\zeta}
\newcommand{\omh}{\hat{\omega}}

\newcommand{\Ch}{M}

\newcommand{\ibest}{\imath_*}
\newcommand{\budget}{n^e}
\newcommand{\budgetV}{\vet{\budget}}
\input{PM_Macros.tex}

\input{macrosFile}

\definecolor{MD}{rgb}{0,0.6,0.2}

\begin{document}

\title{A Multi-Stage Adaptive Sampling Scheme for Passivity Characterization of Large-Scale Macromodels}

\author{\IEEEauthorblockN{Marco De Stefano~\IEEEmembership{Student Member,~IEEE}, Stefano Grivet-Talocia,~\IEEEmembership{Fellow,~IEEE}, Torben Wendt,~\IEEEmembership{Student Member,~IEEE}, Cheng Yang,~\IEEEmembership{Member,~IEEE}, Christian Schuster,~\IEEEmembership{Senior Member,~IEEE}}}

\maketitle

\begin{abstract}
    This paper proposes a hierarchical adaptive sampling scheme for passivity characterization of large-scale linear lumped macromodels. Here, large-scale is intended both in terms of dynamic order and especially number of input/output ports. Standard passivity characterization approaches based on spectral properties of associated Hamiltonian matrices are either inefficient or non-applicable for large-scale models, due to an excessive computational cost. This paper builds on existing adaptive sampling methods and proposes a hybrid multi-stage algorithm that is able to detect the passivity violations with limited computing resources. Results from extensive testing demonstrate a major reduction in computational requirements with respect to competing approaches.
\end{abstract}

\section{Introduction}
	
Macromodeling techniques are generally considered as key enabling factors for the efficient simulation of complex devices and systems. A macromodel approximates the input-output response of a given structure through a compact set of behavioral equations. Such equations are most commonly derived through some fitting process~\cite{tpwrd-1999-VF} applied to a set of sampled responses, usually in the frequency domain. See~\cite{mon-2015-PM} for an overview.

The focus of this work is on macromodeling of passive structures, which are unable to generate energy on their own~\cite{jnl-2007-tadvp-Fundamentals,Willems_dissipative,Wohlers,Anderson}. Model passivity must be therefore checked and enforced through suitable constraints, which are usually applied through a perturbation step applied to an initially non-passive model obtained by some unconstrained fitting~\cite{1262465,4114368,1498835,5715905,6298986,5484629,4492758,4526214,6224200,jnl-2004-cas-Hamiltonian,jnl-2006-tadvp-LargePassivity,jnl-2007-tadvp-AdaptiveSampling,jnl-2008-tadvp-PassivityMethods}. Lack of model passivity can trigger instabilities in transient analyses~\cite{jnl-2009-ijcta-destabilize} and must be therefore avoided with care.

Macromodels are intended to be compact and reduced-order, hence small-size. Nonetheless, the robustness of state of the art macromodeling algorithms has enabled applications that are aggressively driving model complexity beyond current capabilities. One typical example can be full-package modeling for signal, power or coupled signal-power integrity analysis, in which case the number of input/output ports can reach hundreds or even thousands~\cite{hir2019}. A second application is modeling of energy-selective electromagnetic shielding structures~\cite{eleftheriades14_protec_weak_from_stron} formed by conventional enclosures having openings that are closed by regular grids of nonlinear elements~\cite{jnl-2020-temc-torben,Yang2013,wendt19_num_comp_study,yang2016impuls,yang2016design}. An efficient time-domain simulation calls for a macromodel that represents in a compact way the electromagnetic behavior of the shield, and also in this case the number of input/output ports can reach the order of thousands or more. A further complication is provided by the possibly high dynamic order (number of poles) of the macromodels, as required by accuracy requirements over extended frequency bands.

Recent Vector Fitting (VF)~\cite{tpwrd-1999-VF} implementations~\cite{mwcl-2008-Deschrijver-FastVF,jnl-2012-tcpmt-comp} including parallel codes~\cite{jnl-2011-tcpmt-pvf,9123948} have been demonstrated to scale very favorably with the model size. Conversely, standard passivity enforcement schemes are more critical and less scalable, despite the advancements that have been documented over the last two decades~\cite{jnl-2013-tcpmt-parallel}. The major bottleneck is the passivity verification step, which must detect not only whether a given initial model is passive, but should provide also a precise localization of the passivity violations so that a successive perturbation stage can be setup to remove them. Complexity of this model perturbation step is less critical, so passivity enforcement is not considered here. Passivity characterization is far more  demanding in terms of computational resources, both memory and runtime, and therefore more problematic. To the best of Authors' knowledge, there is no documented passivity characterization approach that can provide reliable results in a limited runtime and with limited memory occupation, for systems with input-output ports in the order of hundreds and possibly exceeding one thousand.

The most prominent passivity characterization approaches can be roughly grouped in three main classes, based respectively on Linear Matrix Inequalities (LMI)~\cite{Anderson}, direct sampling, and Hamiltonian matrix properties~\cite{jnl-2004-cas-Hamiltonian,BBK89}. LMIs provide the best approach under a theoretical standpoint, but applicability is limited to small-scale systems. Pure sampling approaches are very fast and provide a potentially good scalability, but they are likely to miss passivity violations making the verification not reliable. Hamiltonian matrices with their spectral properties are the usual method of choice for passivity verification, for small and medium-scale problems. Extension to large-scale systems has been attempted in~\cite{jnl-2006-tadvp-LargePassivity}, possibly through hybridization, pre- or post-processing with adaptive sampling methods~\cite{jnl-2007-tadvp-AdaptiveSampling}. However, the necessary Krylov subspace iterations for the spectral characterization of large and sparse Hamiltonian matrices makes the entire approach infeasible when the number of ports exceeds about one hundred.

The above difficulties motivate the passivity verification approach that we propose in this work. We build on the adaptive sampling scheme of~\cite{jnl-2007-tadvp-AdaptiveSampling}, and we try to avoid completely the use of Hamiltonian matrices and fully coupled 
LMI conditions, which are the main responsible for driving computational cost beyond what is usually affordable in design flows. The main objective that we pursue is to construct an adaptive sampling scheme that is fast and at the same time reliable. We should stress that there is no mathematical guarantee that checking local passivity conditions on a finite number of points will not miss a violation. Therefore, the main objective of this work is to make the presence of undetected passivity violations extremely unlikely. This is the reason why we propose a hybrid scheme that combines different basic sampling methods in a hierarchical way.

The proposed algorithm performs an initial pole-based sampling intended to track fast variations. The results are then used to produce an adaptive frequency warping and subdivision on a possible large number of subbands, which are then subjected to hierarchical refinement~\cite{2016_Dujaili_Suresh_NMSO}. The proposed scheme was tested against a large number of regression test cases, with a number of ports ranging from $P=1$ to $P=640$ and a dynamic order (number of states) ranging from $N=10$ to more than $25\times 10^{3}$. The results show that standard Hamiltonian-based approaches remain competitive and should be preferred for small-scale systems. Conversely, the proposed approach outperforms Hamiltonian tests for medium and large-scale systems, with speedup in runtime reaching and exceeding $200\times$ for the largest cases.

This paper is organized as follows. Section~\ref{sec:background} provides some background information and sets notation. Section~\ref{sec:statement} states the main problem by providing an illustrative example that highlights the main difficulties to be addressed. Section~\ref{sec:freq_warping} introduces the adaptive frequency warping scheme that forms the initial step of proposed sampling scheme. Section~\ref{sec:nmso} introduces the multi-scale search to be applied as a second-stage adaptive refinement. Numerical results are presented in Section~\ref{sec:results} and conclusions are drawn in Section~\ref{sec:conclusions}.

\section{Background and Notation}\label{sec:background}

Throughout this work we consider a lumped Linear Time Invariant (LTI) model, which is known either through a pole-residue expansion or a state-space realization
\begin{equation}\label{eq:model}
    \mat{H}(s) = \sum_{n=1}^{\bar{n}} \frac{\mat{R}_n}{s-p_n} + \mat{R}_0 = \mat{C} (s\eye - \mat{A})^{-1} \mat{B} + \mat{D}
\end{equation}
where $s$ is the Laplace variable and the transfer matrix $\mat{H}(s)\in\Complex^{P\times P}$ with $P$ number of input-output ports. The dynamic order of the model is denoted as $N$, which is the size of the state-space matrix $\mat{A}$. In case the residue matrices $\mat{R}_n$ are full-rank, we have $N= \bar{n}P$, see~\cite{mon-2015-PM,1046889,gilbert1963} for details. In general, $N \gg P$. We assume that the bandwidth where the model is assumed and intended to be accurate is $[0,\omega_{\max}]$, and that all model parameters (poles-residues or state-space matrices) are available from some initial fitting/approximation stage. We assume that all state-space matrices are real-valued and that the model poles $p_n$ (equivalently, the eigenvalues of $\mat{A}$) are strictly stable. This condition is easily enforced by widespread fitting schemes such as VF~\cite{tpwrd-1999-VF}.

\subsection{Passivity verification}\label{sec:pass_check}

We will consider only models~\eqref{eq:model} in scattering representation, so that $\mat{H}(s)$ approximates the scattering matrix of the modeled system. This is not a limitation, since the proposed approach can be extended to other representations (impedance, admittance, or hybrid) with simple modifications. See~\cite{mon-2015-PM} for a complete theoretical framework applicable to all representations. We review below the three main available approaches for passivity verification.

\subsubsection{Sampling local passivity conditions}

For scattering models~\eqref{eq:model} with real-valued realization and strictly stable poles the passivity conditions can be stated as the unique constraint
\begin{equation}\label{eq:sigma_max}
    \sigma_{\max}\{\mat{H}(\jj\omega)\} \leq \gamma = 1 \quad \forall \omega \in \Real.
\end{equation}
This condition expresses that the maximum singular value of the scattering matrix must not exceed the threshold $\gamma=1$ at any real frequency. Under the working assumptions,~\eqref{eq:sigma_max} is a sufficient condition for $\mat{H}(s)$ to be \emph{Bounded Real}, hence passive~\cite{Wohlers,Anderson,jnl-2007-tadvp-Fundamentals}.

Sampling approaches test~\eqref{eq:sigma_max} over a finite number of frequency points $\{\omega_k,\;k=1,\dots,K\}$. Determination of $\sigma_{\max}$ at a single frequency requires $O(P^3)$ operations, hence the overall complexity for $K$ samples amounts to $O(KP^3)$. Since each individual frequency $\omega_k$ can be processed independently, there is no significant memory cost, even for large $P$. The main advantage of this sampling-based check is the availability of all local maxima of $\sigma_{\max}$, obtained as a trivial post-processing. These maxima are essential to setup passivity enforcement through constrained perturbation. The main disadvantage of checking~\eqref{eq:sigma_max} over $K$ finite points is the possibility to miss local maxima exceeding $\gamma=1$ by a small amount or over a small frequency extent, due to the limited number of points that can be processed.

We remark that the (scalar) function of frequency
\begin{equation}\label{eq:pass_metric}
    \sss(\omega) = \sigma_{\max}\{\mat{H}(\jj\omega)\}
\end{equation}
henceforth denoted as \emph{passivity metric} is generally smooth and differentiable, with the possible exception of a finite number of points where two singular values cross. In the latter case, $\sss(\omega)$ is guaranteed to be continuous since all singularties (poles) of $\mat{H}(s)$ fall outside the imaginary axis.

\subsubsection{Linear Matrix Inequalities}

LMI conditions provide a fully algebraic test that does not require sampling. We recall that the scattering system~\eqref{eq:model} is passive (dissipative) if and only if $\exists \mat{P} = \mat{P}^\tran > 0$ such that
\begin{equation}\label{eq:BRL1}
	\begin{pmatrix}
			\mat{A}^\tran \mat{P} + \mat{P} \mat{A} + \mat{C}^\tran \mat{C} & \mat{P} \mat{B} + \mat{C}^\tran \mat{D} \\
		\mat{B}^\tran \mat{P} + \mat{D}^\tran\mat{C} & -({\eye}-\mat{D}^\tran\mat{D})
	\end{pmatrix} \leq 0
	\,.
\end{equation}
This condition, known as Bounded Real Lemma (BRL) or Kalman-Yakubovich-Popov (KYP) Lemma~\cite{Anderson,scherer2000linear}, has the major disadvantage of requiring the determination of the $N\times N$ matrix $\mat{P}$ related to the internal energy storage. The computational cost for checking the feasibility of~\eqref{eq:BRL1} is $O(N^6)$, which can be reduced to $O(N^4)$ with specialized formulations~\cite{KYP-IPM}. This cost becomes exceedingly high for large-scale systems. One additional disadvantage of~\eqref{eq:BRL1} is that this is a pass/fail test which does not provide localization of passivity violations that can be used in a successive enforcement. Passivity enforcement schemes exist that use directly~\eqref{eq:BRL1} as constraint~\cite{1262465,1250052}, but these inherit the poor scalability with large $P$ and $N$, see~\cite{jnl-2008-tadvp-PassivityMethods}.

\subsubsection{Hamiltonian matrices}\label{sec:Ham}

Spectral properties of Hamiltonian matrices provide the standard approach for checking passivity of small and medium-scale models. Under the working assumptions, the model is (strictly) passive if and only if the \emph{Hamiltonian} matrix
\begin{equation}\label{eq:Ham_scat_def}
	\ham{M} =
	\begin{pmatrix}
		\mat{A} + \mat{B}\mat{R}^{-1}\mat{D}^\tran\mat{C} &
		\mat{B}\mat{R}^{-1}\mat{B}^\tran \\
		- \mat{C}^\tran\mat{S}^{-1}\mat{C} & 
		- \mat{A}^\tran - \mat{C}^\tran\mat{D}\mat{R}^{-1}\mat{B}^\tran
	\end{pmatrix}\,.
\end{equation}
with $\mat{R}={\eye}-\mat{D}^\tran\mat{D}$ and $\mat{S}={\eye}-\mat{D}\mat{D}^\tran$
has no purely imaginary eigenvalues $\mu_k = \jj\hat{\omega}_k$. Such eigenvalues denote the crossing points of any singular value of $\mat{H}(\jj\omega)$ with the passivity threshold $\gamma=1$, see~\cite{jnl-2004-cas-Hamiltonian,BBK89}. Therefore, they provide also localization of passivity violations. Unfortunately, the eigenvalue spectrum of $\ham{M}$ is required, whose numerical determination scales as $O((2N)^3)$. Extension to large-scale and sparse (decompositions) was documented in~\cite{jnl-2006-tadvp-LargePassivity,jnl-2007-tadvp-AdaptiveSampling}, allowing some improvement for systems with moderate $P$ and possibly large $\bar{n}$. Extension to large port count $P$ appears problematic, since the Krylov subspace iterations required for detection of purely imaginary eigenvalues of $\ham{M}$ require the repeated inversion of full $P \times P$ matrices, leading to a cost that once again becomes impractical in the large-scale case.

We remark that~\eqref{eq:Ham_scat_def} is applicable only when state-space matrix $\mat{D}$ has no singular values too close to $\gamma=1$. In this case, it is preferable to process the \emph{extended Hamiltonian pencil} $(\ham{M}_{\rm e},\ham{K})$ defined as~\cite{5456981}
\begin{equation}
    \ham{M}_{\rm e} =
    \begin{bmatrix}
		\mat{A} & \mat{0} & \mat{B} & \mat{0}\\
		\mat{0} & -\mat{A}^\tran & \mat{0} & -\mat{C}^\tran \\
		\mat{0} & \mat{B}^\tran & -{\eye} & \mat{D}^\tran \\
		\mat{C} & \mat{0} & \mat{D} & -{\eye}
	\end{bmatrix} \quad
    \ham{K} =
	\begin{bmatrix}
		{\eye} & \mat{0} & \mat{0} & \mat{0} \\
		\mat{0} & {\eye} & \mat{0} & \mat{0} \\
		\mat{0} & \mat{0} & \mat{0} & \mat{0} \\
		\mat{0} & \mat{0} & \mat{0} & \mat{0}
	\end{bmatrix}
	\label{eq:M_eigv_ext}	
\end{equation}
to find its purely imaginary generalized eigenvalues $\mu_k$. Some of the test cases that will be investigated in Section~\ref{sec:results} fall in this situation, in which the standard Hamiltonian matrix~\eqref{eq:Ham_scat_def} is ill-defined. Throughout this work, we use~\eqref{eq:M_eigv_ext} only when $|\sigma_{\max}\{\mat{D}\}-1|<10^{-4}$, since the associated computational cost is higher than for~\eqref{eq:Ham_scat_def}.

\subsection{Discussion}

\begin{table}[t]
    \centering
    \caption{Asymptotic scaling factors of computational costs for different passivity characterization approaches}
    \label{tab:cost}
    \begin{tabular}{ll}
         Characterization method & Operations \\ 
         \hline
         Linear Matrix Inequalities~\cite{KYP-IPM} & $O(N^4)$ \\ 
         Hamiltonian (full) & $O((2N)^3)$ \\ 
         Sampling & $O(K(P^3+P^2\bar{n}))$ \\ 
         \hline
         $N$: model order (number of states)\\
         $P$: number of input-output ports\\
         $K$: number of frequency samples\\
         $\bar{n}$: number of (common) poles
    \end{tabular}
\end{table}

Table~\ref{tab:cost} summarizes the number of elementary operations for the three leading passivity characterization approaches. Only the leading terms are included, and the number of floating point operations is reported up to a constant factor which depends on the specific algorithm or numerical library used to implement each method.

As an example, let us consider a system with $P=400$ ports and $\bar{n}=50$ poles (hence $N=20000$). The LMI and the Hamiltonian approaches would require about 3~GB and 12~GB of storage, respectively, with a relative CPU cost with respect to sampling a huge number $K=10^5$ of frequencies about $2\cdot 10^4 \times$ and $9\times$, respectively. From this example and Table~\ref{tab:cost}, it is evident that the only framework that has the potential to scale favorably with the problem size is the sampling approach. Therefore, we concentrate our efforts on the sampling scheme, with the main objective of reducing the likelihood of missing passivity violations and classifying of a non-passive model as passive.

\section{Problem statement and goals}\label{sec:statement}

\begin{figure}[t]
    \centerline{\includegraphics[width=\columnwidth]{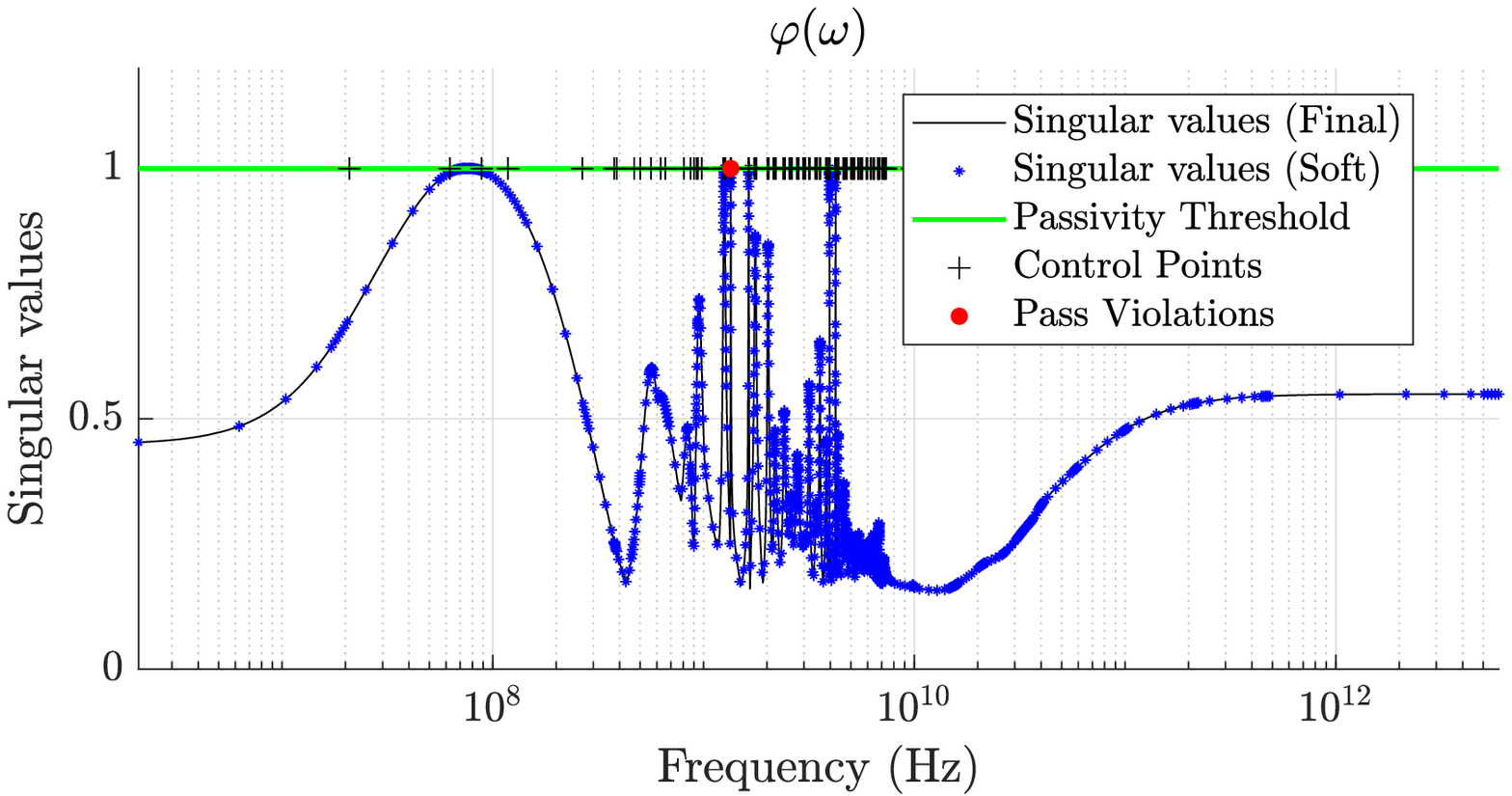}}
    \vspace{4mm}
    \centerline{\includegraphics[width=\columnwidth]{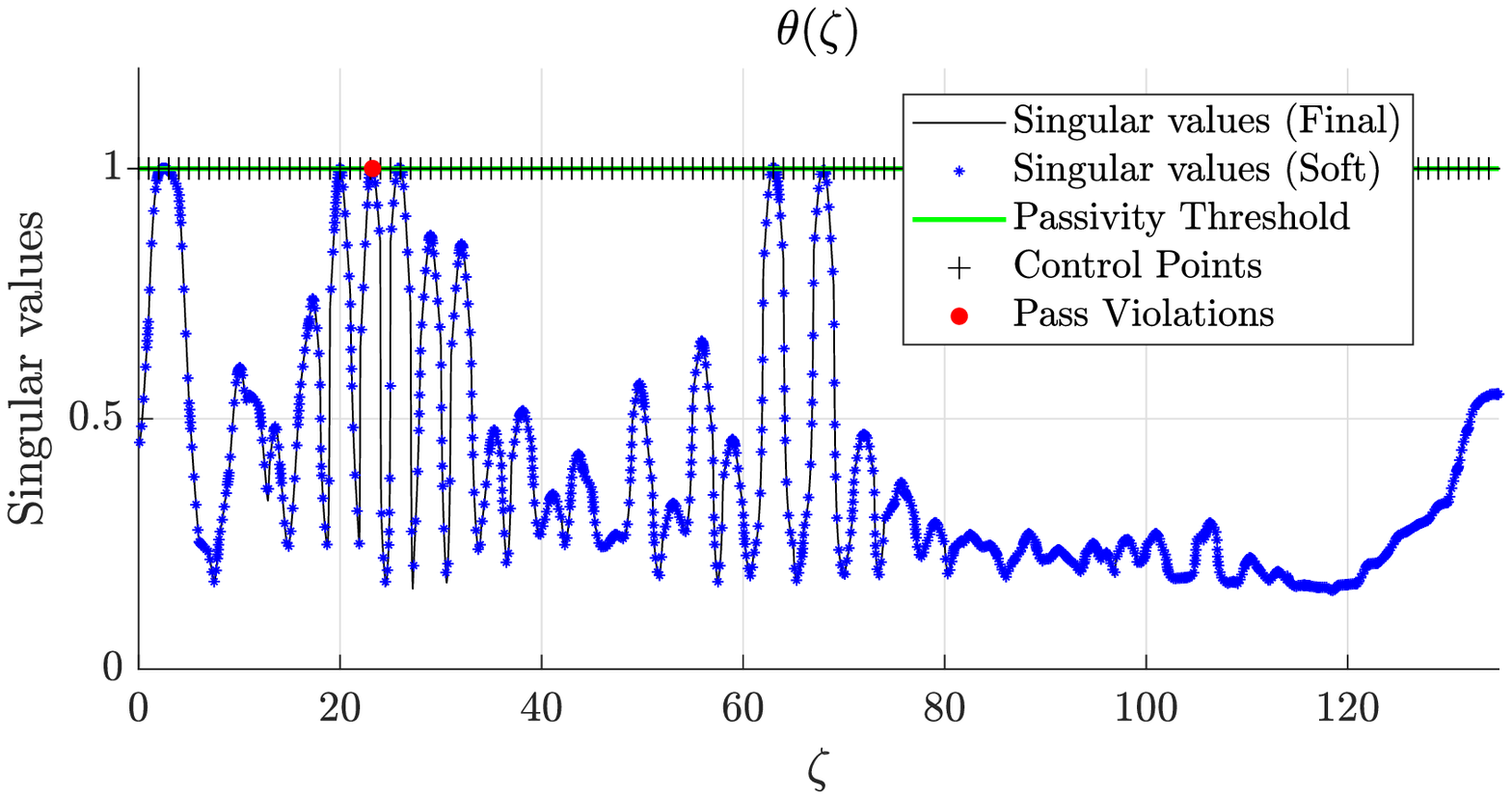}}
    \caption{Top panel: graphical illustration of the adaptive sampling-based passivity characterization. Bottom panel: rescaled passivity metric $\ssr(\omw)$ after adaptive frequency warping.}
    \label{fig:adaptive_example}
\end{figure}

\begin{figure}[t]
    \centerline{\includegraphics[width=\columnwidth]{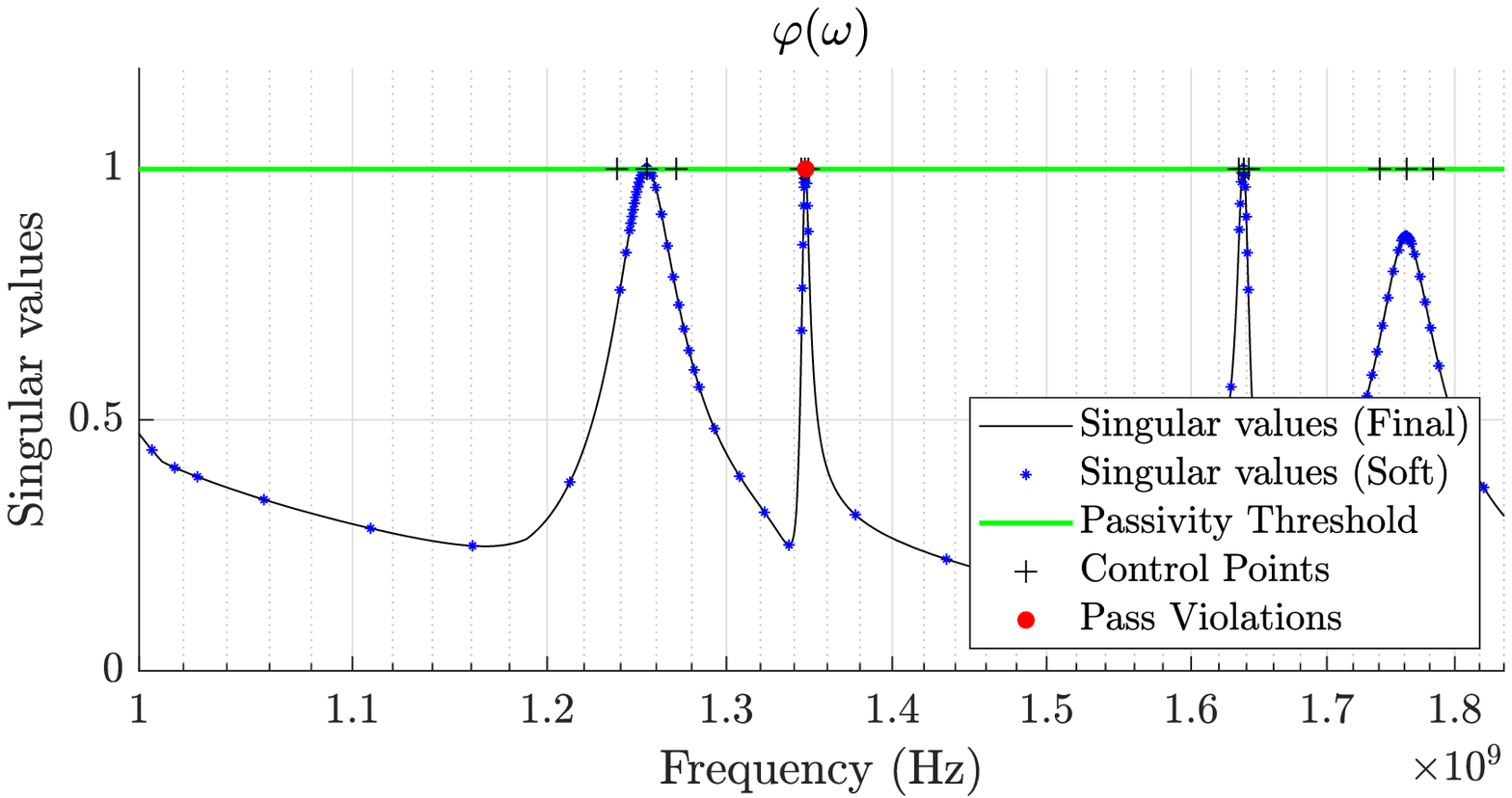}}
    \vspace{4mm}
    \centerline{\includegraphics[width=\columnwidth]{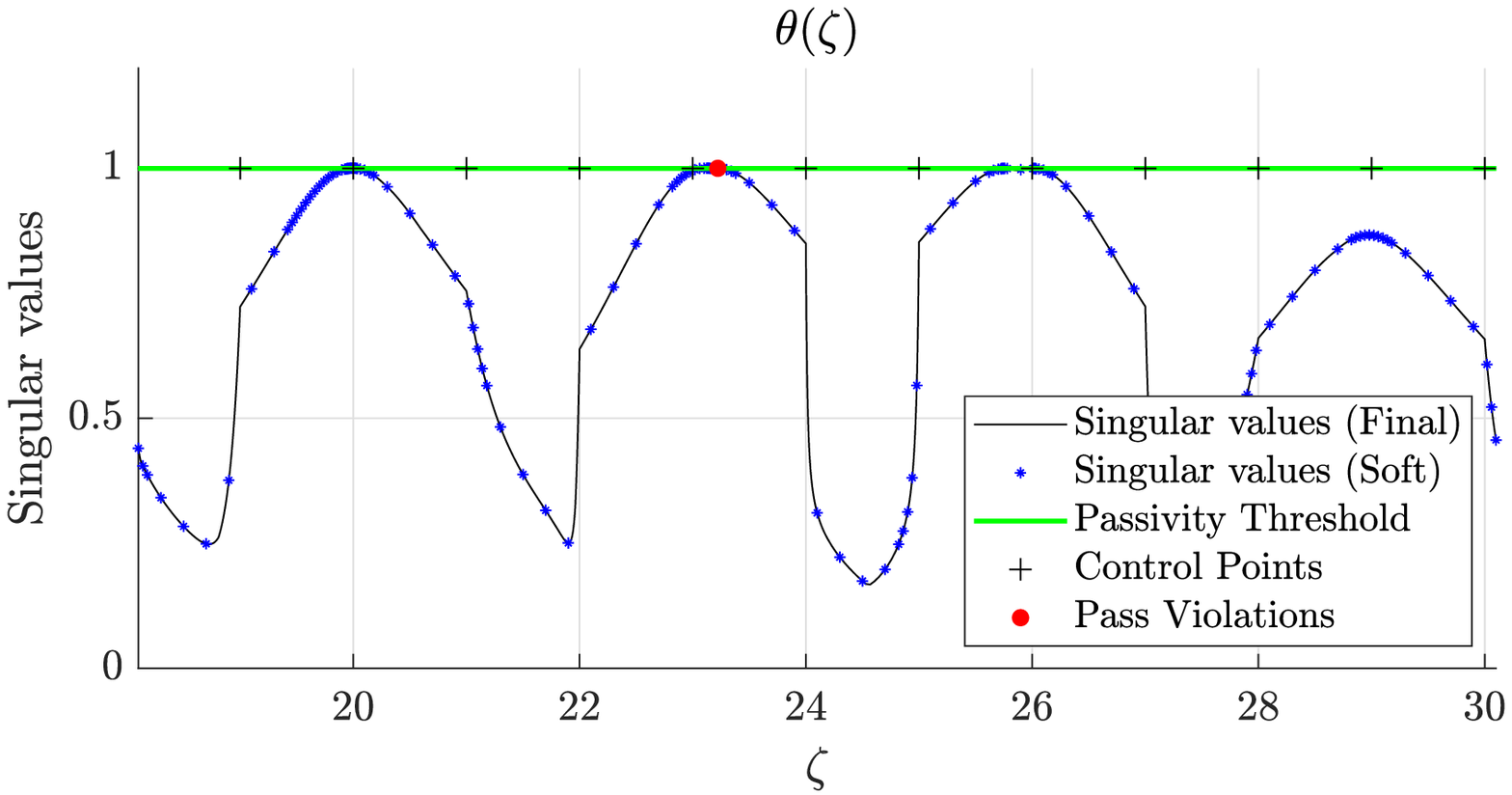}}
    \caption{Enlarged view of Figure~\ref{fig:adaptive_example}.}
    \label{fig:adaptive_example_zoom}
\end{figure}

Considering the passivity condition~\eqref{eq:sigma_max} and the definition of the passivity metric~\eqref{eq:pass_metric}, we can state the main problem that is here addressed as: \emph{find all local maxima larger than a given threshold of a continuous univariate function of frequency over the entire frequency axis.} This problem may appear simple, but various difficulties arise in practical implementation. These are discussed below with reference to the top panel of Fig.~\ref{fig:adaptive_example} and its enlarged view in the top panel of Fig.~\ref{fig:adaptive_example_zoom}.

\begin{itemize}
    \item The passivity metric $\sss(\omega)$ is usually characterized by multiple peaks, and it is necessary to identify all of them in order to check if they exceed the threshold $\gamma$. It is imperative not to miss any local maximum.
    \item Some peaks can be extremely narrow, hence they can be almost invisible with an inadequate sampling.
    \item Some peaks may be characterized by a local maximum that is very close to the threshold; in such cases, it may be difficult to detect whether there exists a passivity violation with $\sss(\omega)>\gamma$ at some frequency point.
\end{itemize}

The above difficulties can only be addressed by adapting the sampling process, in order to automatically refine the resolution where $\sss(\omega)$ is characterized by fast variations and/or its values are close to the threshold $\gamma$. At the same time, it is mandatory to limit as much as possible the total number $K$ of computed samples.

These objectives are here pursued through a multi-stage adaptive sampling scheme. First, a preprocessing step is applied to split the frequency band into a possibly large number of subbands, whose individual extent is adaptively defined and rescaled so that the total variation of the passivity metric becomes nearly uniform (see Section~\ref{sec:freq_warping}). Second, a hierarchical tree-based refinement is applied within each individual subband to track all local maxima (see Section~\ref{sec:nmso}). The combination of these two stages will be demonstrated to outperform competing approaches, still retaining (and in some cases even improving) the reliability of algebraic passivity verification methods.


\section{Step 1: pole-based adaptive frequency warping}\label{sec:freq_warping}

A preview of the results of this first step is available in the bottom panels of Fig.~\ref{fig:adaptive_example} and its enlarged view of Fig.~\ref{fig:adaptive_example_zoom}. We want to identify a nonlinear (invertible) frequency transformation $\omw={\cal W}(\omega)$ that induces the following change of variable
\begin{equation}
    \ssr(\omw) = \sss({\cal W}^{-1}(\omw)) 
\end{equation}
with the main objective of ``flattening'' the local variations of $\sss(\omega)$, so that the resulting $\ssr(\omw)$ is characterized by peaks with an approximately uniform width.

The change of variable is constructed based on a set of control points $\Omega = \{\omh_\ell,\;\ell=0,\dots,L\}$ such that
\begin{equation}\label{eq:control_points}
    0 = \omh_0 < \omh_1 < \dots < \omh_\ell < \omh_{\ell+1} < \dots < \omh_L = \infty
\end{equation}
These points are used to build a local linear map between the normalized subband $\omw\in[\ell,\ell+1]$ and the corresponding subband $\omega\in[\omh_\ell, \omh_{\ell+1}]$. The global change of variable ${\cal W}$ is constructed piecewise as
\begin{equation}
    \omw = \ell + \frac{\omega -\omh_\ell }{\Delta_{\ell}}\quad 
    \forall \omega \in [\omh_\ell, \omh_{\ell+1}], \quad \ell=0,\dots,L-2
\end{equation}
where $\Delta_\ell = \omh_{\ell+1} - \omh_{\ell}$ is the extent of the $\ell$-th subband. The last (infinite-sized) subband is handled as
\begin{equation}
    \omw = \ell + \frac{\omega -\omh_\ell}{\omega}\quad \forall \omega \in [\omh_\ell, \omh_{\ell+1}], \quad \ell = L-1
\end{equation}

The piecewise linear frequency warping function ${\cal W}$ maps the entire frequency axis $\omega\in[0,+\infty)$ into the normalized interval $[0,L]$. The effects are visible by comparing top and bottom panels of Fig.~\ref{fig:adaptive_example} or Fig.~\ref{fig:adaptive_example_zoom} (enlarged view). The control points highlighted with crosses in the top panels are mapped to uniformly distributed points in the bottom panels. Wherever the density of the control points is higher, corresponding to fast variations of $\sss(\omega)$, the final effect after renormalization will be to stretch narrow peaks and shrink wide peaks, so that all local maxima in the normalized domain $\omw$ are located at peaks with a comparable and statistically uniform width. We see from Fig.~\ref{fig:adaptive_example_zoom} that the number of local maxima in each subband is at most few units (usually only one), with a most probable situation characterized by this maximum occurring at one edge of the subinterval. This favorable situation is guaranteed by a careful selection of the control points, discussed below.

\subsection{Choosing control points}\label{sec:ctrl_points}

The strategy for constructing the control points $\omh_{\ell}$ aims at reducing $\Delta_{\ell}$ in those regions that are characterized by fast variations of $\sss(\omega)$, and conversely relaxing and enlarging $\Delta_{\ell}$ where $\sss(\omega)$ is slowly-varying or nearly constant. The main underlying assumption is that variations of the singular values are induced by variations in the transfer matrix elements, which in turn are induced by the location of each pole/residue term. Therefore, we start with the pole-based sample distribution discussed in~\cite{jnl-2007-tadvp-AdaptiveSampling}, which for each pair of complex conjugate poles $p_n=\alpha_n \pm \jj \beta_n$ contributes the following candidate frequency samples
\begin{equation}\label{eq:pole_based_sampling}
    \omega_{n,r} = \beta_n + \alpha_n \text{tan}\frac{r\pi}{2(R+1)} \quad r=-R,\cdots,R
\end{equation}
The parameter $R$ is used to control how many samples are due to each pole term in a partial fraction expansion of the model response. Only frequencies $\omega_{n,r}\geq 0$ are retained and assembled in a single set $\Omega$, that is then sorted in ascending order. A final scan is performed to remove samples that are closer than a predefined resolution $\Delta\omega$, which is defined as in~\cite{jnl-2007-tadvp-AdaptiveSampling} as
\begin{equation}\label{eq:Delta_omega}
    \Delta\omega = \frac{p_{\max}}{N\rho}
\end{equation}
where $\rho\gg 1$ is another control parameter and $p_{\max} = \max(\omega_{\max},\max_n |p_n|)$.

The above-defined set $\Omega$ was tested and proved to be inadequate for a pure sampling-based passivity check. In particular,
\begin{enumerate}
    \item the parameters setting suggested in~\cite{jnl-2007-tadvp-AdaptiveSampling} results in a large set of subbands, improving the robustness of the procedure but reducing its overall speed. A more efficient check would call for small $R$. This strategy would provide an adequate sample distribution in band, but the passivity metric variations for $\omega>\omega_{\max}$ might not be sampled appropriately.
    \item variations induced by real poles are not adequately sampled by~\eqref{eq:pole_based_sampling}.
    \item poles that are very close to the imaginary axis lead to subbands with very small $\Delta_\ell$, whose edges may be removed accidentally by~\eqref{eq:Delta_omega}, leading to inaccurate sampling.
\end{enumerate}
The above issues~1 and~2 are here solved by introducing three separate control parameters $R_\nu$ with $\nu=\{{\rm cp},{\rm rp}, {\rm hf}\}$ to be used respectively for in-band complex pole pairs, real poles, and poles with either real or imaginary part magnitude that is close to or larger than $\omega_{\max}$. In particular,
\begin{itemize}
    \item $R_{\rm cp}$ can be safely chosen very small, even $R_{\rm cp}=1$;
    \item $R_{\rm rp}$ should be larger, with at least $R_{\rm rp}\geq 2$;
    \item $R_{\rm hf}$ should also be larger, with at least $R_{\rm hf}\geq 3$.
\end{itemize}
Issue~3 is partially compensated by modifying $\omega_{n,r}$ corresponding to highly resonant poles in~\eqref{eq:pole_based_sampling} by setting
\begin{equation}
    \alpha_n \leftarrow c\cdot \alpha_n \quad \textrm{if} \quad Q_n \approx \frac{\abs{\beta_n}}{2 \abs{\alpha_n}} > Q_{\max} 
\end{equation}
where $Q_{\max}$ is here set to $500$ and $c=50$. Finally, since the entire real line $\omega\in\Real$ needs to be sampled, we add $\omega=\infty$ together with the following $\kappa+1$ logarithmically spaced samples extending $d$ decades beyond the fitting model band
\begin{equation}
      \omega_\nu = \omega_{\max} \cdot 10^{d \frac{\nu}{\kappa}}\quad \nu= 0,\cdots, \kappa
\end{equation}
where $d$ is an additional control parameter. Typical values are $d\geq 0.5$ and $\kappa \geq 2$.

In summary, frequency warping and renormalization is here achieved by a pole-based control point distribution, whose behavior can be controlled by the eight parameters summarized in Table~\ref{table:parametersWarping}. The above guidelines and parameter settings have been heuristically proven to be quasi-optimal over a large number of test cases, see Section~\ref{sec:results}. It should be considered that, when inserted in a passivity enforcement loop, the number of subbands can be adaptively increased as the passivity violations are iteratively removed and become more difficult to identify, simply by increasing the values of $R_\nu$, as discussed and illustrated in Sec.~\ref{sec:params} through an application example

\section{Step 2: A Modified Naive Multi-scale Search Optimization}\label{sec:nmso}

The second step of proposed passivity verification algorithm applies a hierarchical adaptive sampling process to the rescaled passivity metric $\ssr(\omw)$ within each normalized subband $\omw\in\mathcal{X}_\ell = [\ell,\ell+1]$, with the main objective of identifying all local maxima that exceed the passivity threshold $\gamma=1$.

Among the several adaptive algorithms offered by the literature, our starting point is the so-called Naive Multi-scale Search Optimization (NMSO) algorithm~\cite{2016_Dujaili_Suresh_NMSO}, which is based on a tree-search divide-and-conquer strategy. Although based on adaptive tree refinement, the NMSO algorithm may suffer from the presence of many local maxima, since its main objective is to find the unique global maximum in its search domain. This is the main reason why we performed the subband splitting and frequency warping in Section~\ref{sec:freq_warping}. Each subband $\mathcal{X}_\ell=[\omw_\ell,\omw_{\ell+1}]$ is constructed to include a very limited number of local maxima, usually only one. Therefore, we propose a repeated and independent application of a suitably constructed NMSO scheme to each subband $\mathcal{X}_\ell$ independently, so that the identification of all local maxima is greatly simplified. The discussion below focuses on $\mathcal{X}=\mathcal{X}_0=[0,1]$ without loss of generality.

\begin{figure}[t]
	\centerline{\includegraphics[width=\columnwidth]{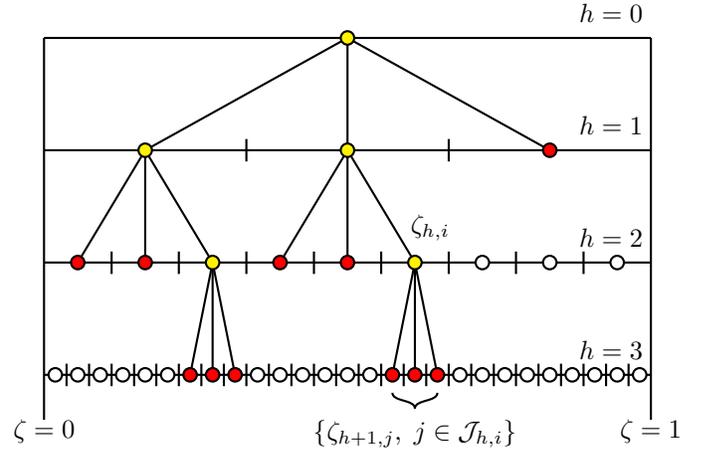}}
	\vspace*{0.2cm}
	\caption{Graphical illustration of a $M$-tree with $M=3$. All leaves up to level $h=3$ are denoted by dots. Filled dots represent leaves that have been evaluated (set $\mathcal{E}^h$). Red dots denote all leaves currently in the tree, whereas yellow dots are nodes that have been expanded and relabeled ($M$ odd).}
	\label{fig:Tree}
\end{figure}

The proposed algorithm builds a tree $\mathcal{T}$ whose nodes are iteratively refined at multiple scales $h\geq 0$, in order to partition the search space $\mathcal{X}$ at any level $h$ into a set of $\Ch^{h}$ cells
\begin{equation}
    \mathcal{X}_{h,i}=[i \Ch^{-h},(i+1)\Ch^{-h}], \quad i=0,\dots, \Ch^{h}-1.
\end{equation}
Each cell has a center point $\omw_{h,i} = (i+\frac{1}{2})\Ch^{-h}$, denoted as \emph{leaf} or \emph{node} in the following. The partition factor $\Ch\geq2$ defines the number of children to be expanded from each leaf of the tree when increasing refinement level. The main objective is to find all local maxima of $\ssr(\omw)$ over the search space $\mathcal{X}$, by evaluating $\ssr_{h,i} = \ssr(\omw_{h,i})$ at a minimal number of tree leaves, which are determined adaptively through hierarchical refinement of the tree. Figure~\ref{fig:Tree} provides a graphical illustration, where the entire set $\mathcal{L}^h$ of leaves of a complete tree at a given maximum refinement level $h$ is represented by dots, while the set $\mathcal{E}^h \subseteq\mathcal{L}^h$ of evaluated leaves is depicted by filled dots.

\subsection{Adaptive sampling rules}

The various assumptions and features of proposed scheme are illustrated below, whereas a pseudocode version is available in Algorithm~\ref{al:PDNMSO}.

\begin{algorithm}[t]
\caption{Find $\ssr_{\max} = \arg\max \ssr(\omw)$ for $\omw\in[0,1]$}
\label{al:PDNMSO}
\begin{algorithmic}[1]
\REQUIRE $h_0$, $M$, $\delta\omw$, $\delta\ssr$, $\delta\eta$, $\varepsilon$, $\varrho$, $\budget$, $\delta\budget$
\STATE Initialize $\mu=0$, $h\leftarrow h_0$, $\mathcal{B}_{0}=\emptyset$
\STATE Initialize $\mathcal{C}_0 = \{\omw_{h,i},\;i=0,..,M^h-1\}$, $K=M^h$
\STATE Initialize $\ssr_{\max} = -\infty$
\WHILE{$\mathcal{C}_\mu \neq \emptyset$}
    \STATE Choose current point $\omw_{h,\ibest}$ via~\eqref{eq:choose_best}
    \STATE Expand current point via~\eqref{eq:refinement}
    \STATE $K\leftarrow K+M-1$
    \STATE Update $\ssr_{\max}\leftarrow \max\{\ssr_{\max},\{\omw_{h+1,j},\;j\in\mathcal{J}_{h,\ibest}\}\}$
    \IF{$K>\budget$}
        \IF{$\mathcal{U}_1 \; \texttt{AND}\;  (\mathcal{U}_2\; \texttt{OR}\; \mathcal{U}_3)$}
            \STATE update budget via~\eqref{eq:budget_update}
            \STATE update $\varepsilon \leftarrow \varrho\, \varepsilon$
        \ELSE
            \STATE go to~\ref{al:return}
        \ENDIF
    \ENDIF
    \IF{$\mathcal{S}_1 \; \texttt{OR}\;  \mathcal{S}_2\;  \texttt{OR}\; \mathcal{S}_3$}
        \STATE insert new points in basket via~\eqref{eq:new_points_in_B}
        \STATE $h\leftarrow h_{\min}$
        \STATE reset $\varepsilon$
    \ELSE
        \STATE flag new points for refinement via~\eqref{eq:new_points_in_C}
        \STATE $h\leftarrow h+1$
    \ENDIF
    \STATE $\mu \leftarrow \mu+1$
\ENDWHILE
\RETURN samples $\mathcal{E}_{\mu} = \mathcal{C}_\mu \cup \mathcal{B}_\mu$ and maximum $\ssr_{\max}$ \label{al:return}
\end{algorithmic}
\end{algorithm}

\subsubsection{General structure}

The dynamic evolution of the tree is performed through successive iterations with index $\mu$. At any iteration, the set of evaluated nodes $\mathcal{E}_\mu$ is characterized by three levels
\begin{itemize}
    \item $h_{\min}$: the minimum level of any node in $\mathcal{E}_\mu$
    \item $h_{\max}$: the maximum level of any node in $\mathcal{E}_\mu$
    \item $h$: the level of the node being processed, also denoted as \emph{current node}; $h$ is henceforth denoted as \emph{current level}.
\end{itemize}
The elements of $\mathcal{E}_{\mu}$ are split into two subsets
\begin{equation}
    \mathcal{E}_{\mu} = \mathcal{C}_{\mu} \cup \mathcal{B}_{\mu}
\end{equation}
where $\mathcal{C}_{\mu}$ collects candidates for refinement at the next iteration, and $\mathcal{B}_{\mu}$ forms the so-called \emph{basket}, which includes those nodes that do not need to be refined since a stop condition is verified (see below).

\subsubsection{Constraints}

We aim at minimizing the number of function evaluations. Therefore, we allow a total initial \emph{budget} $\budget$ of function evaluations. \junk{Depending on tree evolution, this budget may be updated dynamically through iterations, see Algorithm~\ref{al:budget}. Therefore, the algorithm is parameterized by a sequence of tentative budgets collected in vector $\budgetV$. If the initial budget needs to be increased, the next element in vector $\budgetV$ will be used as current budget.}
Another constraint on which we build our implementation is to enforce $M$ to be odd.

\subsubsection{Initialization}

The algorithm is initialized for $\mu=0$ by choosing an initial refinement level $h_0\geq 0$ and evaluating the passivity metric $\ssr$ (our target function) at the corresponding nodes $\{\omw_{h_0,i},\;i=0,\dots,\Ch^h-1\}$. Such samples are inserted into $\mathcal{E}_0 = \mathcal{C}_0$, with the initial basket $\mathcal{B}_0=\emptyset$.

\subsubsection{Expansion (refinement)}

The expansion process splits a given cell $\mathcal{X}_{h,i}$ into its $M$ children at level $h+1$, denoted as
\begin{equation}
    \mathcal{X}_{h+1,j}, \quad  j\in {\cal J}_{h,i}=\{Mi,\dots,M(i+1)-1\}
\end{equation}
The set ${\cal J}_{h,i}$ collects the indices of the nodes within \emph{cone of influence} of node $\omw_{h,i}$ under refinement by one level. We remark that using an odd partition factor $M$ guarantees that the midpoint of the central children cell coincides with that of the expanded leaf
\begin{equation}
    \omw_{h+1,Mi+\lfloor M/2 \rfloor} = \omw_{h,i},
\end{equation}
which does not need to be re-evaluated but simply relabeled, see Fig.~\ref{fig:Tree}. Note also that elimination of node $\omw_{h,i}$ upon its refinement may trigger an update of $h_{\min}$ as a consequence of this relabeling. Refinement of one leaf $\omw_{h,i}$ can thus be expressed as
\begin{equation}\label{eq:refinement}
    \mathcal{E}_{\mu+1} = \left(\mathcal{E}_{\mu} - \{\omw_{h,i}\}\right)  \cup
    \{\omw_{h+1,j}:\; j\in {\cal J}_{h,i}\}
\end{equation}

\subsubsection{Choosing leaves for refinement}

One key operation is the selection of best node for refinement among a set of candidates in $\mathcal{C}_{\mu}$. This is simply the leaf corresponding to the largest value of the passivity metric
\begin{equation}\label{eq:choose_best}
    \omw_{h,\ibest} = \arg\max \{\ssr(\omw_{h,i}):\, \omw_{h,i}\in\mathcal{C}_{\mu}\} 
\end{equation}
selected among all leaves at the current level $h$. This choice is motivated by the main objective of finding local maxima.

\subsubsection{Stopping conditions, classification, and restarts}

The proposed implementation checks the nodes generated upon refinement~\eqref{eq:refinement} to determine if they need to be further refined. This is achieved through various stopping conditions
\begin{itemize}
    \item condition $\mathcal{S}_1$: a target resolution has been reached. Considering current level $h$ being refined into nodes at levels $h+1$ this condition is expressed as
    \begin{equation}
        M^{-h-1} < \delta\omw
    \end{equation}
    where $\delta\omw$ is a control parameter.
    \item condition $\mathcal{S}_2$: the largest variation among adjacent refined nodes is within a prescribed tolerance. Denoting
    \begin{equation}
        \Delta_{h,i} = \max\{ |\ssr_{h+1,j+1} - \ssr_{h+1,j} |,\; j=Mi,\dots,M(i+1)-2 \}
    \end{equation}
    condition $\mathcal{S}_2$ can be expressed as
    \begin{equation}
        \Delta_{h,i} < \delta\ssr
    \end{equation}
    where $\delta\ssr$ is a control parameter.
    \item condition $\mathcal{S}_3$: the largest variation among adjacent refined nodes is less than their distance from the passivity threshold
    \begin{equation}
        \Delta_{h,i} < | \widehat{\ssr}_{h,i} - \gamma |,
    \end{equation}
    where
    \begin{equation}
        \widehat{\ssr}_{h,i} = \max \{\ssr(\omw_{h+1,j}),\; j\in {\cal{J}}_{h,i} \}
    \end{equation}
    This latter condition is only considered when a minimum resolution has been achieved
    \begin{equation}\label{eq:minres_check}
        M^{-h-1} < \delta\eta
    \end{equation}
    where $\delta\eta$ is control parameter.
\end{itemize}
After a refinement process, if all conditions $\mathcal{S}_1$, $\mathcal{S}_2$, $\mathcal{S}_3$ are false, all nodes $\omw_{h+1,j}$ obtained from refinement are flagged as potentially critical and will be candidates for refinement at the next iteration, so that
\begin{equation}\label{eq:new_points_in_C}
    \mathcal{C}_{\mu+1} = \left(\mathcal{C}_\mu - \{\omw_{h,i} \} \right) \cup
    \{\omw_{h+1,j},\; j\in {\cal{J}}_{h,i} \}
\end{equation}
At the same time, current level is increased as
\begin{equation}
    h \leftarrow h+1
\end{equation}
Otherwise, the new nodes $\omw_{h+1,j}$ from refinement are flagged as non-critical and are inserted in the basket
\begin{equation}\label{eq:new_points_in_B}
    \mathcal{B}_{\mu+1} = \left(\mathcal{B}_\mu - \{\omw_{h,i} \} \right) \cup
    \{\omw_{h+1,j},\; j\in {\cal{J}}_{h,i} \}.
\end{equation}
These nodes will not be further refined. The latter condition will trigger a \emph{restart} by resetting current level to
\begin{equation}
    h \leftarrow h_{\min}
\end{equation}
This condition enhances the exploration capabilities of the search.

\subsubsection{Budget updates}

In principle, when the budget of function evaluations is reached upon a refinement step, the algorithm should stop. However, it may be possible that some critical regions to be explored or refined are still present. 
Therefore, a budget update (increase) is triggered by the three conditions below
\begin{itemize}
    \item condition $\mathcal{U}_1$: all nodes from last refinement are detected as passive
    \begin{equation}
        \widehat{\ssr}_{h,i} < \gamma
    \end{equation}
    \item condition $\mathcal{U}_2$: at least one node from last refinement is closer to the passivity threshold than a prescribed relative threshold $\varepsilon$
    \begin{equation}
        \frac{\gamma - \widehat{\ssr}_{h,i}}{\widehat{\ssr}_{h,i}} < \varepsilon
    \end{equation}
    \item condition $\mathcal{U}_3$: the largest variation among adjacent refined nodes exceeds their smallest distance from the passivity threshold
    \begin{equation}
        |\gamma - \widehat{\ssr}_{h,i}| < \Delta_{h,i}
    \end{equation}
    As for condition $\mathcal{S}_3$, this latter condition is only checked when a minimum resolution has been achieved according to~\eqref{eq:minres_check}.
\end{itemize}
When $\mathcal{U}_1$ holds and one among $\mathcal{U}_2$, $\mathcal{U}_3$ is verified, budget update is achieved by setting
\begin{equation}\label{eq:budget_update}
    \budget \leftarrow \budget + \delta\budget
\end{equation}
where $\delta\budget$ is a control parameter that can even be iteration-dependent. In order to preserve the effectiveness of condition $\mathcal{U}_2$ when the tree level $h$ increases, any budget update~\eqref{eq:budget_update} triggers a reduction of $\varepsilon$ by a factor $\varrho<1$ as $\varepsilon\leftarrow\varrho\varepsilon$. The original value of $\varepsilon$ is restored upon restarts.

\subsubsection{Basket reuse}

The above-listed strategies prevent reusing and further refining the leaves in the basket $\mathcal{B}_\mu$, which is appropriate when execution speed is to be privileged with respect to accuracy. Nevertheless, the procedure can be easily modified to include the elements of $\mathcal{B}_\mu$ in the set of candidates for future expansions $\mathcal{C}_{\mu +1}$, as in the original NMSO scheme~\cite{2016_Dujaili_Suresh_NMSO}. This \emph{basket reuse} strategy should be used for more aggressive sampling, such as required for model qualification at the end of passivity enforcement loops (see below).

\subsection{Implementation: choosing algorithm parameters}\label{sec:params}

The proposed two-step adaptive sampling algorithm is controlled by two sets of parameters for step-1 and step-2. These parameters are collected in Tables~\ref{table:parametersWarping} and~\ref{table:parametersNMSO}, respectively. Both tables report our suggestion for three possible execution modes. Settings labeled as \emph{soft} are adequate for passivity checks that are executed at the beginning of a passivity enforcement loop, where only the largest passivity violations are of interest and an aggressive accuracy is not necessary. Conversely, settings labeled as \emph{hard} are adequate when results are expected to be very accurate, and passivity violations are very small, such as at the last iteration(s) of passivity enforcement loops. Settings labeled as \emph{final} are adequate for final model qualification.

\input{paramTable}

\begin{figure}
    \centerline{\includegraphics[width=\columnwidth]{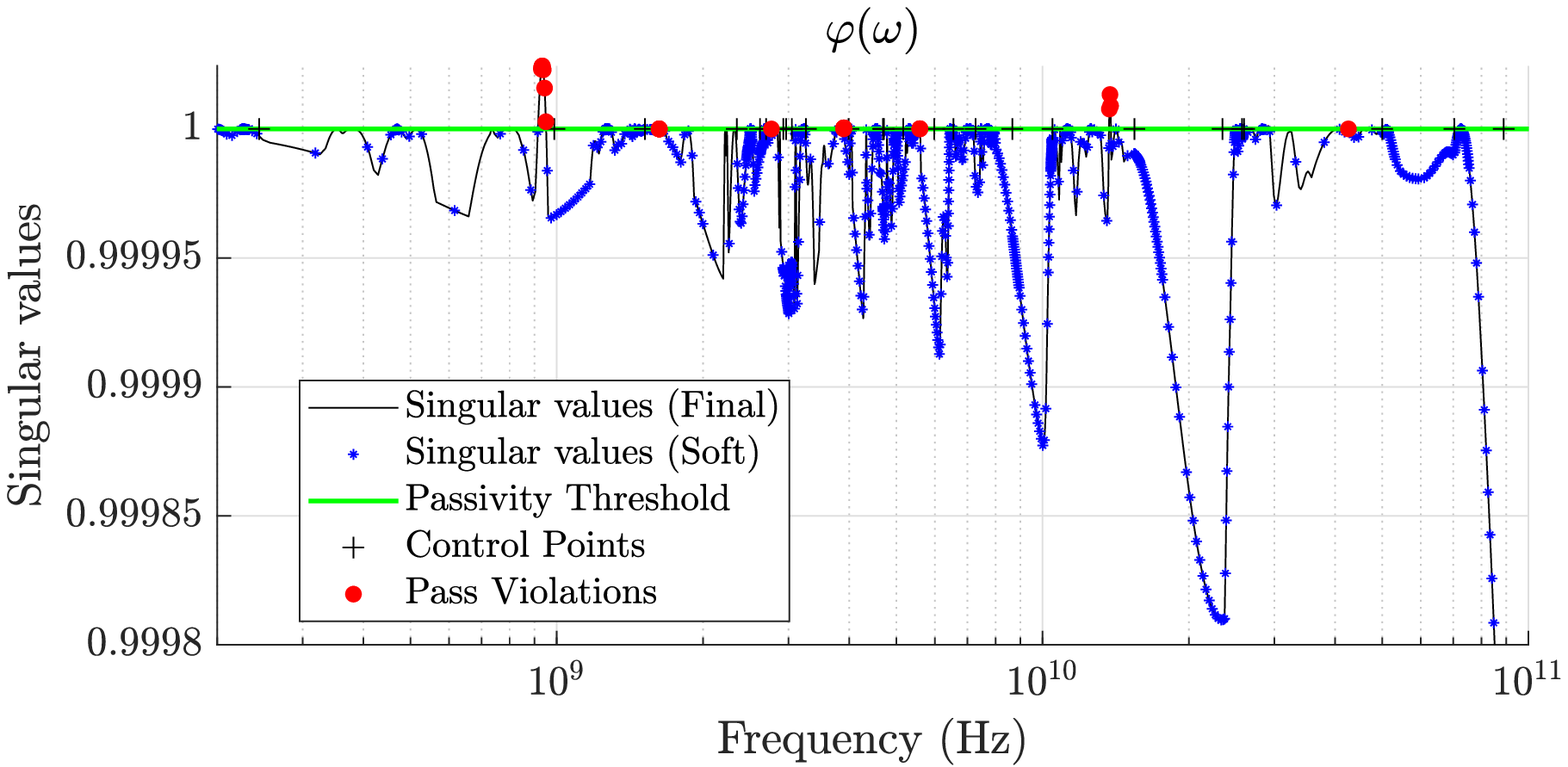}}
    \vspace{4mm}
    \centerline{\includegraphics[width=\columnwidth]{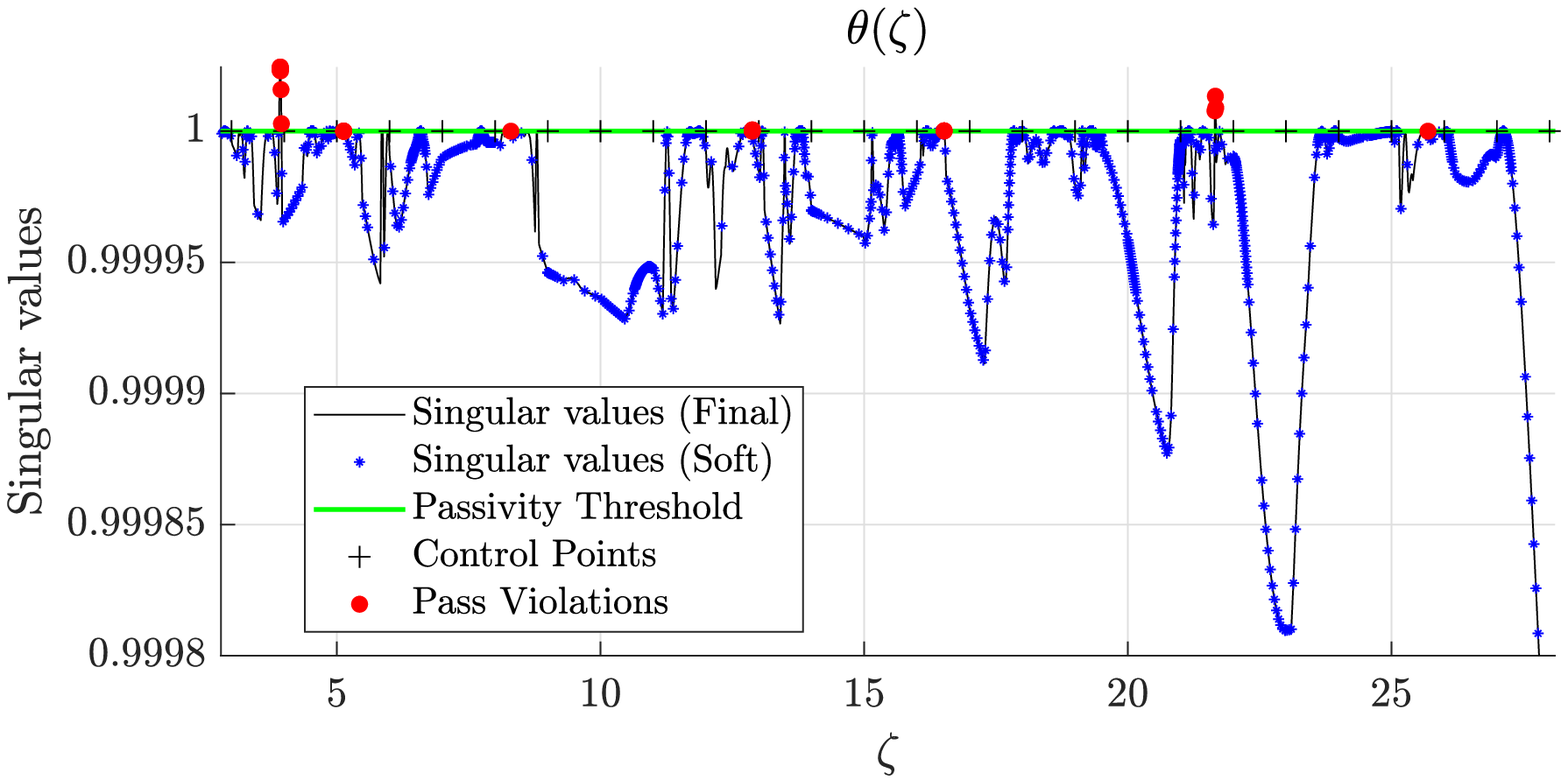}}
    \caption{Test case \#444. Top panel: graphical illustration of the adaptive sampling-based passivity characterization. Bottom panel: rescaled passivity metric $\ssr(\omw)$ after adaptive frequency warping.}
    \label{fig:adaptive_example2}
\end{figure}

The differences between the two \emph{soft} and \emph{final} settings in algorithm performance are illustrated through an example. Figure~\ref{fig:adaptive_example2} compares the results obtained by both settings on a representative test case, precisely \#444 with reference to the complete benchmark database analyzed below in Sec.~\ref{sec:447}. The structure has $P=92$ ports, with a model having $\bar{n}=16$ pole-residue terms and a corresponding state-space matrix size $N=1472$.

\begin{figure}
    \centerline{\includegraphics[width=\columnwidth]{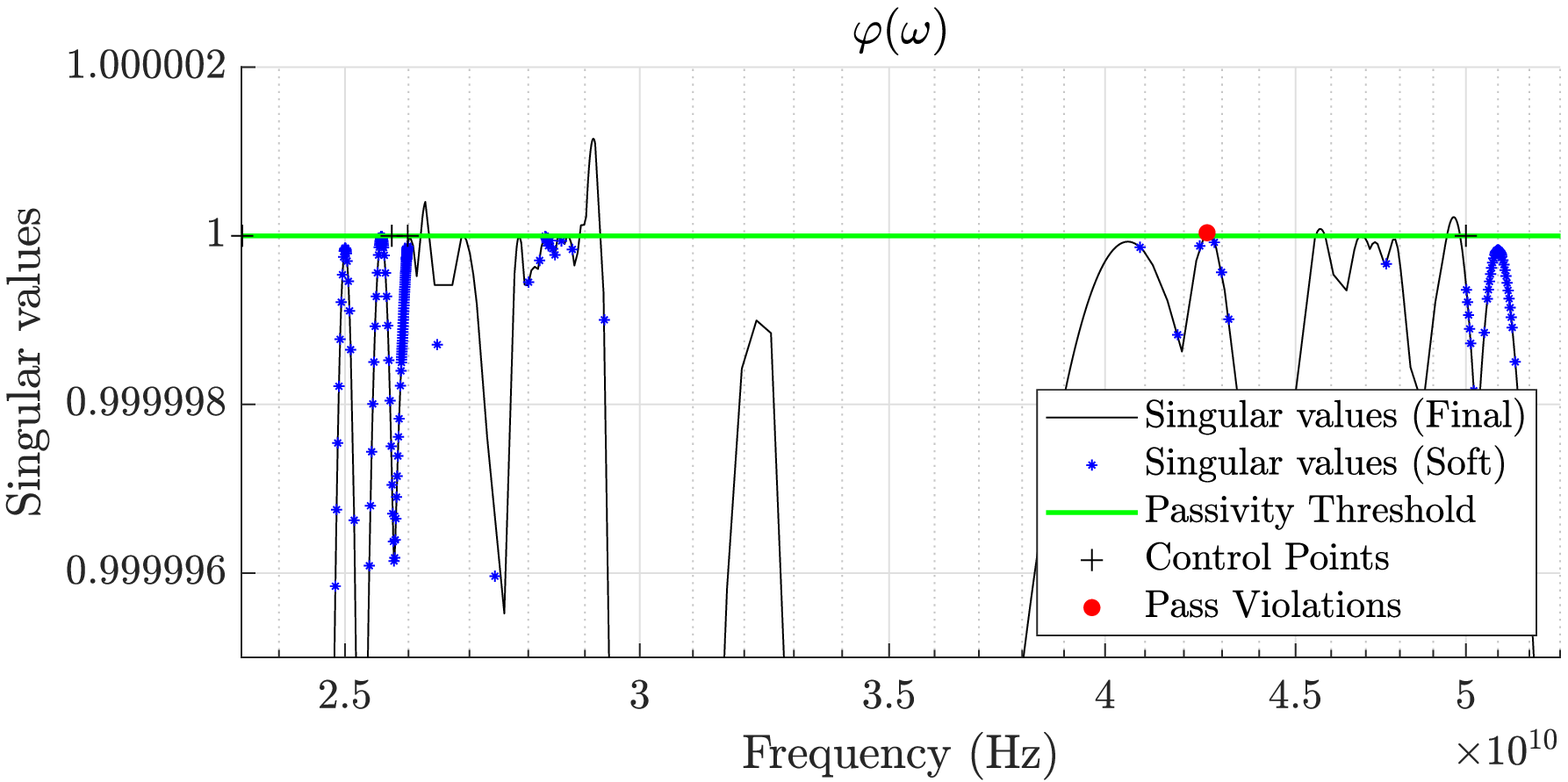}}
    \vspace{4mm}
    \centerline{\includegraphics[width=\columnwidth]{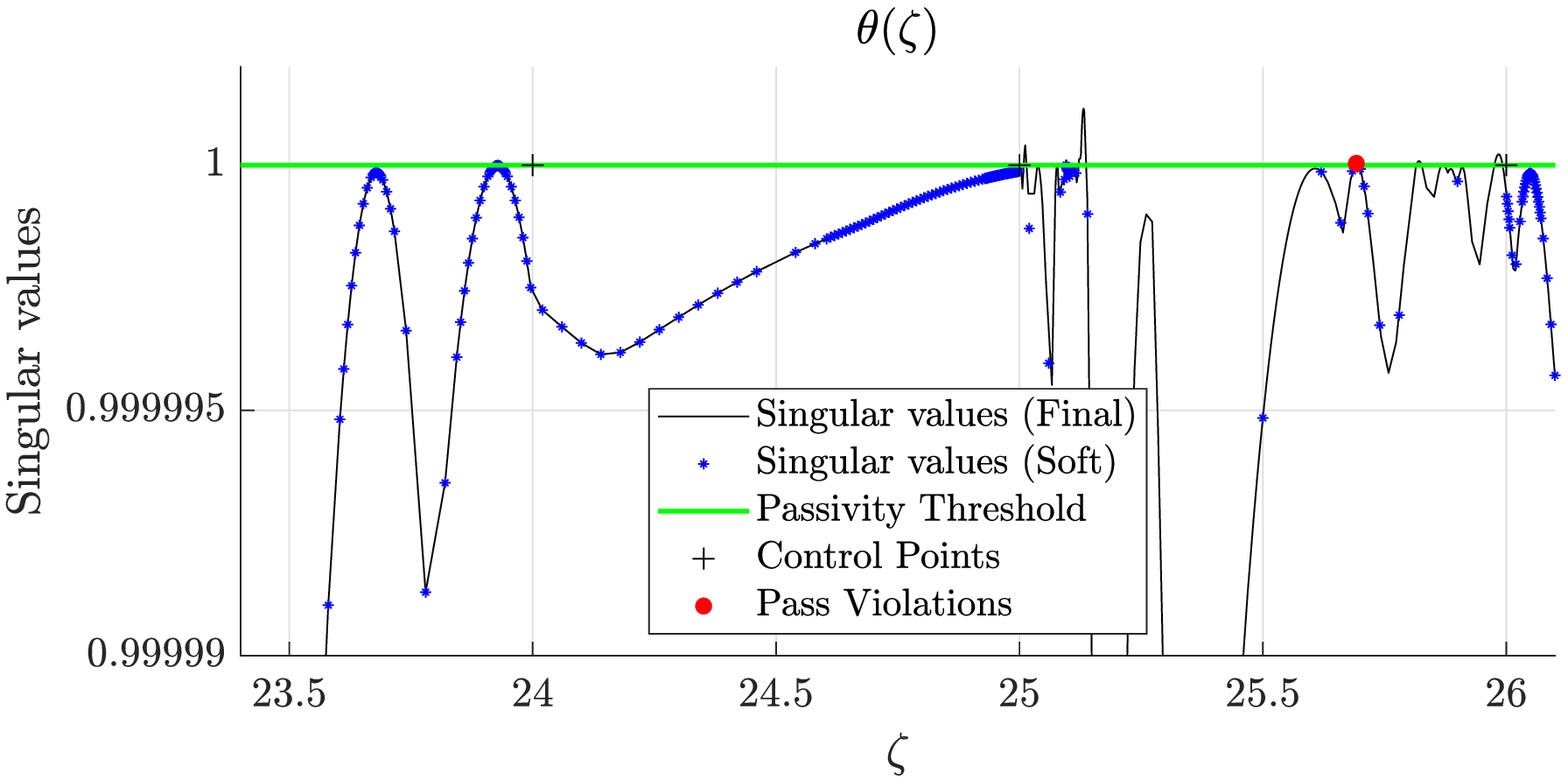}}
    \caption{Zoom of Fig.~\ref{fig:adaptive_example2} on a frequency band with small passivity violations.}
    \label{fig:adaptive_example_zoom2}
\end{figure}

The proposed adaptive sampling check in \emph{soft} mode resulted in $L=30$ control points and $K=2255$ evaluated frequency samples, $28$ of which denoting passivity violations with $\sss(\omega)>\gamma$. 
Algorithm in \emph{hard} mode requested $K=4245$ samples with $L=55$ and $47$ violation points, whereas \emph{final} mode led to $K=9827$ samples evaluation, with $L=55$ and $271$ violation points. It should be noted that the above detected violation points can be at the edges of each subband $\mathcal{X}_\ell$, in which case they should be dropped since not corresponding to actual local maxima over the entire frequency axis. After suitable postprocessing, the actual number of detected local maxima largest than the passivity threshold resulted 
$7$, $14$, and $21$ after \emph{soft}, \emph{hard}, and \emph{final} mode checks, respectively. These results are consistent with the asymptotic outcome of the algorithm used in \emph{soft} mode, imposing $n^e = \infty$.
Figure~\ref{fig:adaptive_example2} and its enlarged view of Fig.~\ref{fig:adaptive_example_zoom2} compare results of \emph{soft} and \emph{final} mode, showing that the latter provides a more accurate detection of even very small violations over narrow frequency bands.

\section{Numerical Results}\label{sec:results}

We document now the performance of proposed algorithm on a large set of test cases. Section~\ref{sec:447} presents the results of a systematic testing campaign on a database of interconnect macromodels, while Sections~\ref{sec:box} and~\ref{sec:via_array} concentrate on two representative large-scale application examples. All documented numerical results have been obtained on a Workstation based on Core i9-7900X CPU running at 3.3~GHz with 64~GB RAM.

\subsection{Consistency and performance}\label{sec:447}

\begin{figure*}[t]
\centerline{\includegraphics[width=\textwidth]{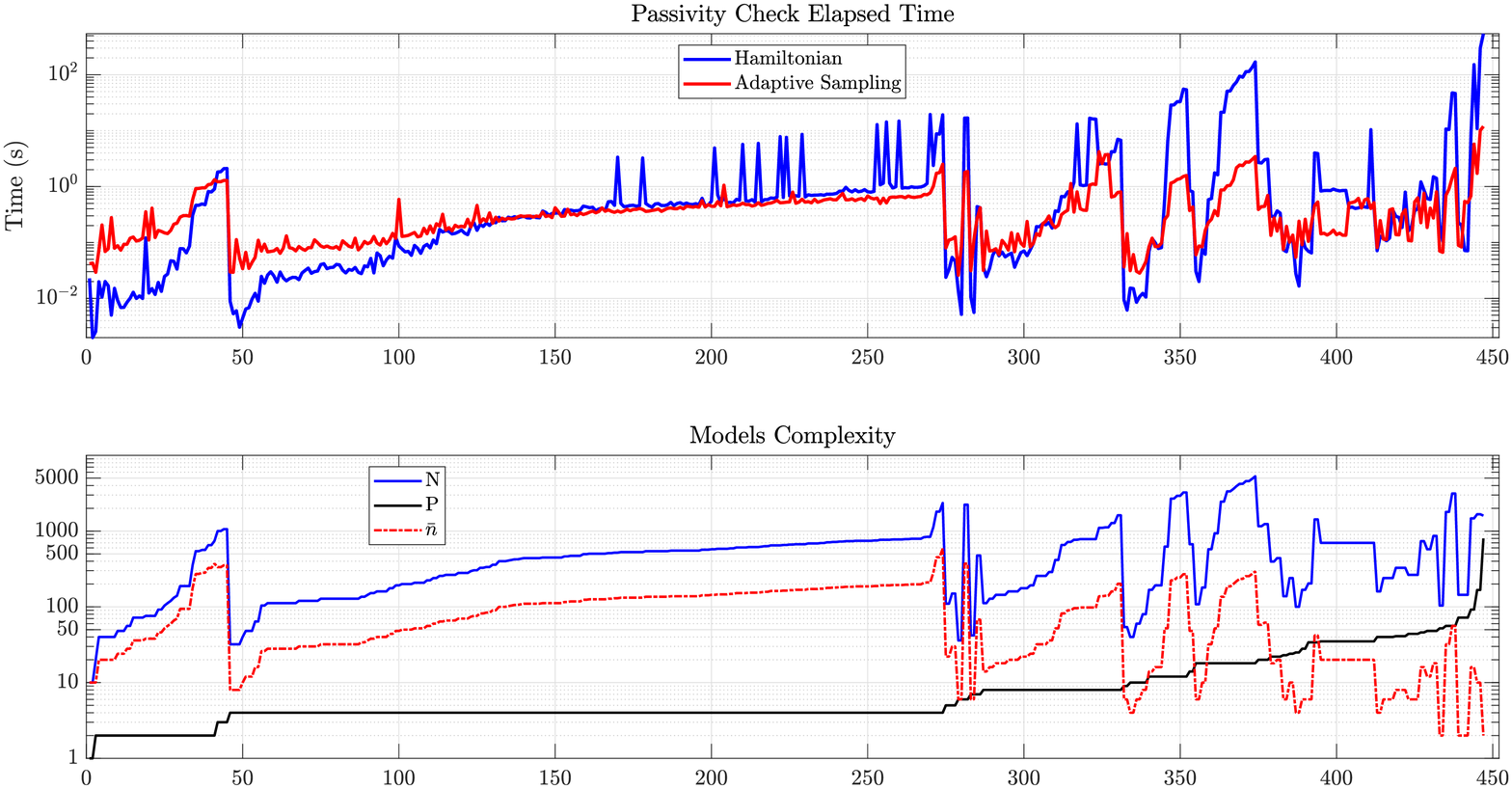}}
\caption{Time performances and models complexity for all the investigated 447 test cases.}
\label{fig:time_and_complexity}
\end{figure*}

The proposed passivity check algorithm was applied to a large set of benchmarks, including 447 models of different size and complexity (details in the bottom panel of Fig.~\ref{fig:time_and_complexity}). A total of 243 models form a database constructed by applying a standard VF scheme as available in the commercial software IdEM~\cite{idem}. The remaining examples were obtained by applying few passivity enforcement iterations on some models from the first set. Since the passivity metric changes completely through passivity enforcement iterations, these latter models can be considered as independent. Moreover, since passivity enforcement iteratively removes passivity violations, models that are processed by few enforcement iterations are characterized by smaller violations, which are therefore potentially more difficult to be detected.

\begin{table}[b]
\caption{Summary of passivity check results on all benchmark cases. Models are classified as: True Positive (\textbf{TP}), False Positive (\textbf{FP}), False Negative (\textbf{FN}), see text}
\label{table: Summary_20201006}
\centering
\begin{tabular}{|c|c|c|c|c|c|}
\hline
\# Tests & Passive & TP & FP & FN & Passive but FN \\
\hline
447 & 179 & 446 & 0 & 1 & 1\\
\hline
\end{tabular}
\end{table}

\begin{figure}
    \centerline{\includegraphics[width=\columnwidth]{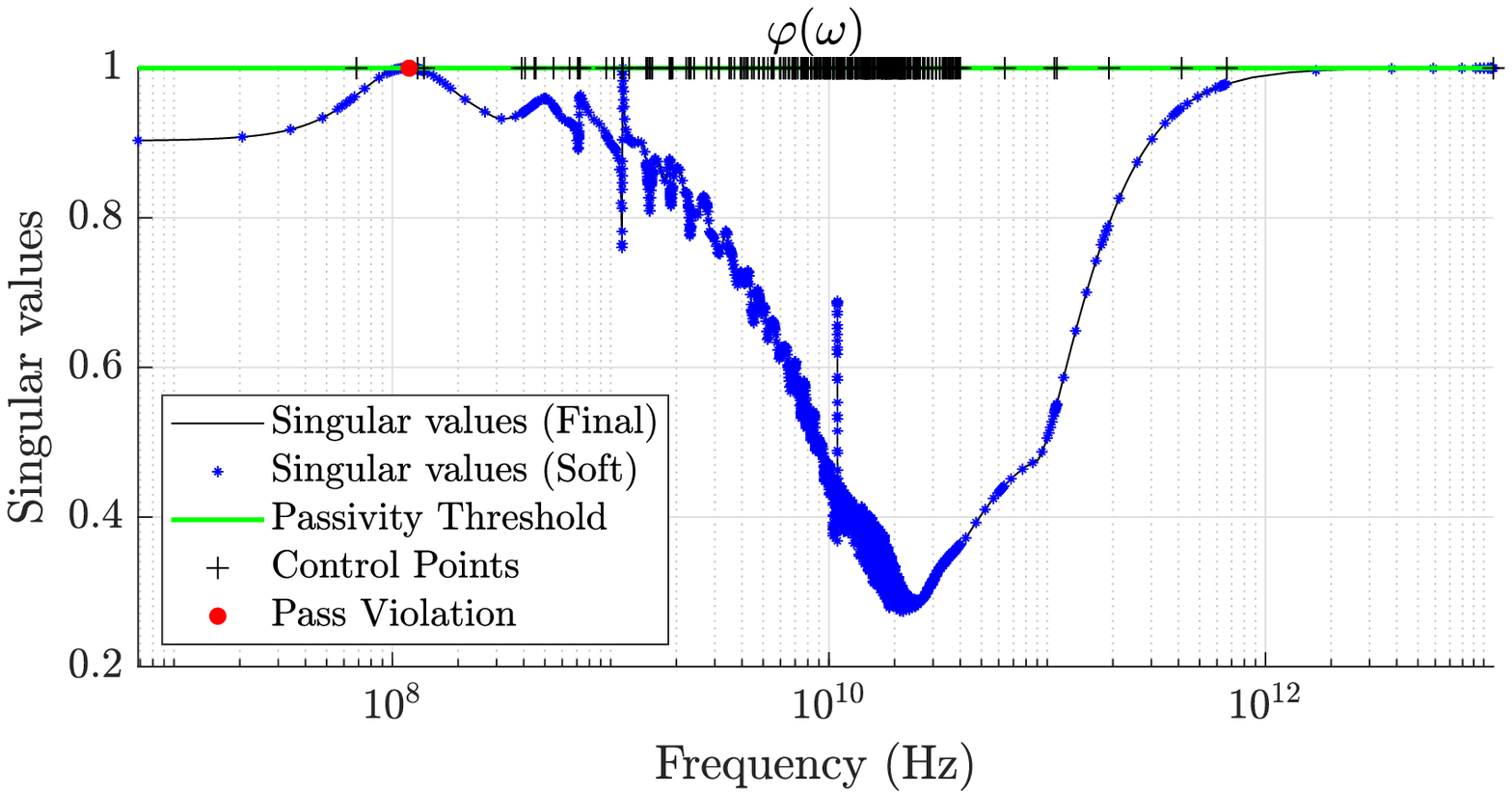}}
    \vspace*{4mm}
    \centerline{\includegraphics[width=\columnwidth]{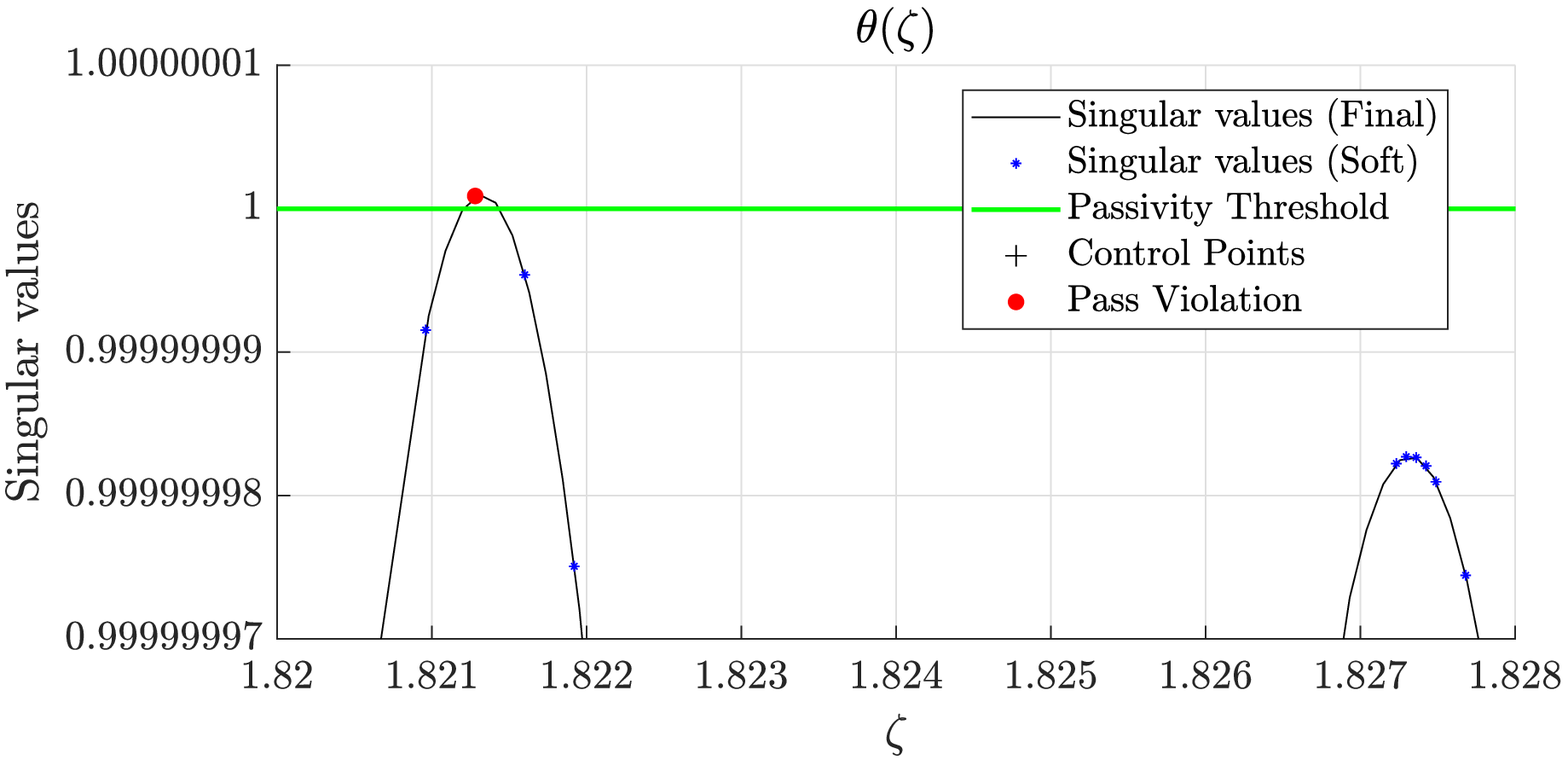}}
    \caption{Passivity characterization of the unique False Negative case (\#260), showing a residual and marginal passivity violation that is not detected by the Hamiltonian test. Bottom panel: zoom on the passivity metric $\ssr(\omw)$ after frequency warping.}
    \label{fig: Mod260_FN}
\end{figure}

A summary of the results is shown in Table~\ref{table: Summary_20201006}, where the passivity characterization offered by proposed scheme is compared to a state-of-the-art Hamiltonian-based check based on~\eqref{eq:Ham_scat_def} or~\eqref{eq:M_eigv_ext}, see Sec.~\ref{sec:Ham}. The various models are classified as
\begin{itemize}
    \item True Positive (\textbf{TP}), if both proposed adaptive sampling and the Hamiltonian based passivity check were in agreement, producing the same outcome (passive or not passive); 
    \item False Positive (\textbf{FP}), if the adaptive sampling approach identified a passive model while the Hamiltonian check did not;
    \item False Negative (\textbf{FN}), if the Hamiltonian check classified a model as passive while the adaptive sampling did not;
    \item the last column \textbf{Passive but FN} indicates models that are really non-passive even if the Hamiltonian check did not find any passivity violation.
\end{itemize}
We see that the proposed algorithm agreed in the 99.9\% of the cases with the Hamiltonian check. The proposed algorithm gave only one different result with respect to the Hamiltonian check, providing a False Negative result. It turns out that this model was actually not passive (with a very tight passivity violation region, and $|\sigma_1-1|\approx 1\cdot10^{-10}$), but imaginary Hamiltonian eigenvalues were not detected. Figure~\ref{fig: Mod260_FN} provides a graphical illustration. We conclude from this systematic testing campaign that proposed scheme may result even more accurate than Hamiltonian algebraic tests, which appear to be more sensitive and prone to ill-conditioning in extreme cases.

The time performances of the two passivity check strategies are presented in the top panel of Fig.~\ref{fig:time_and_complexity}. In all cases a MATLAB-based single-thread prototypal code was used, using a full Hamiltonian eigensolver. We see that the Hamiltonian check remains the method of choice for small-scale models (number of ports and poles is reported in the bottom panel for comparison). For small-scale models, both Hamiltonian and proposed check produce their results in a fraction of a second. When it comes to large-scale models, the situation is drastically different. Proposed scheme has a runtime that for all investigated cases is below 10 seconds, which can be as much as $50\times$ faster than the Hamiltonian check.

\subsection{A large-scale multiport shielding enclosure}\label{sec:box}

\begin{figure}
    \centerline{\includegraphics[width=0.8\columnwidth]{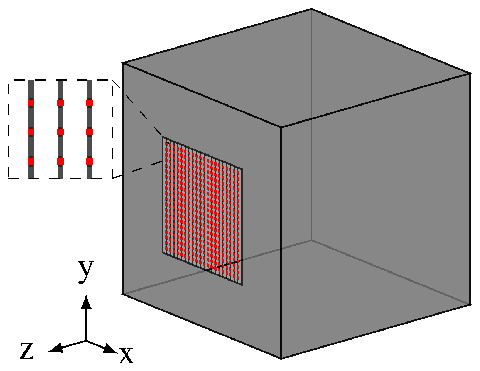}}
    \caption{A perfectly conducting box enclosure designed for energy-selective shielding. Size: $50 \times 50 \times 50$~cm. Red dots depict lumped ports distributed on a regular $20\times 20$ grid throughout the $25 \times 25$~cm aperture.}
    \label{fig:box}
\end{figure}

\begin{figure}
    \centerline{\includegraphics[width=\columnwidth]{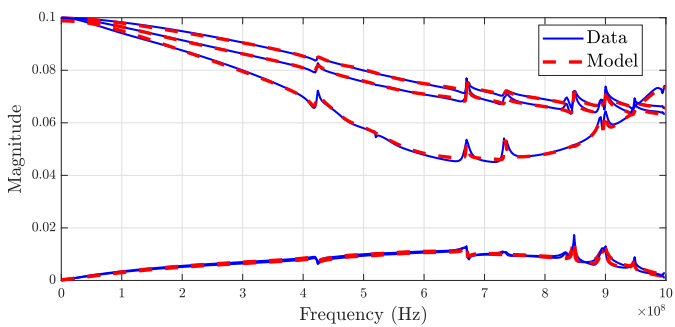}}
    \caption{Model vs data comparison of selected responses of the shielding enclosure.}
    \label{fig:box_responses}
\end{figure}

\begin{figure}
    \centerline{\includegraphics[width=\columnwidth]{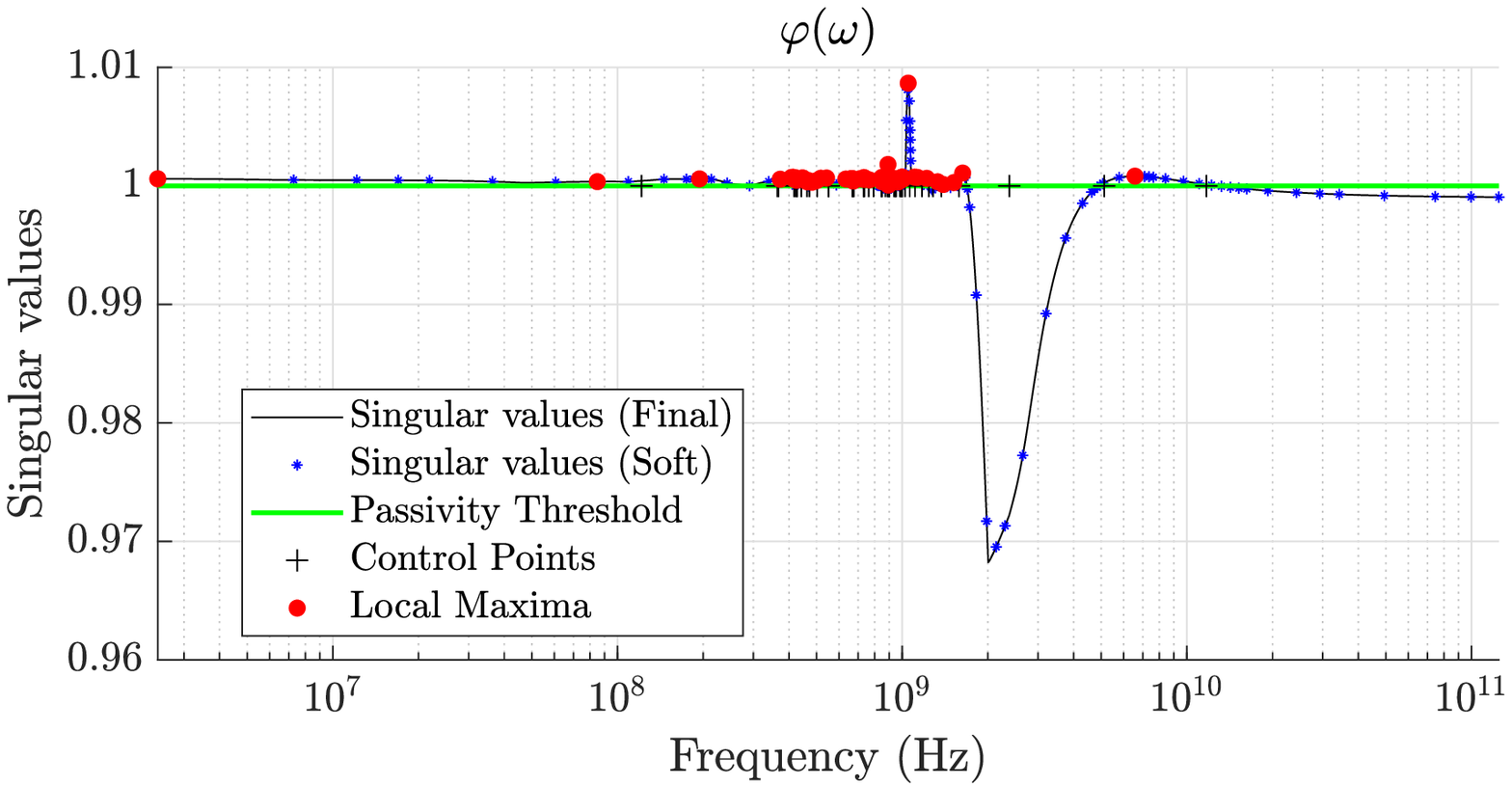}}
    \vspace*{4mm}
    \centerline{\includegraphics[width=\columnwidth]{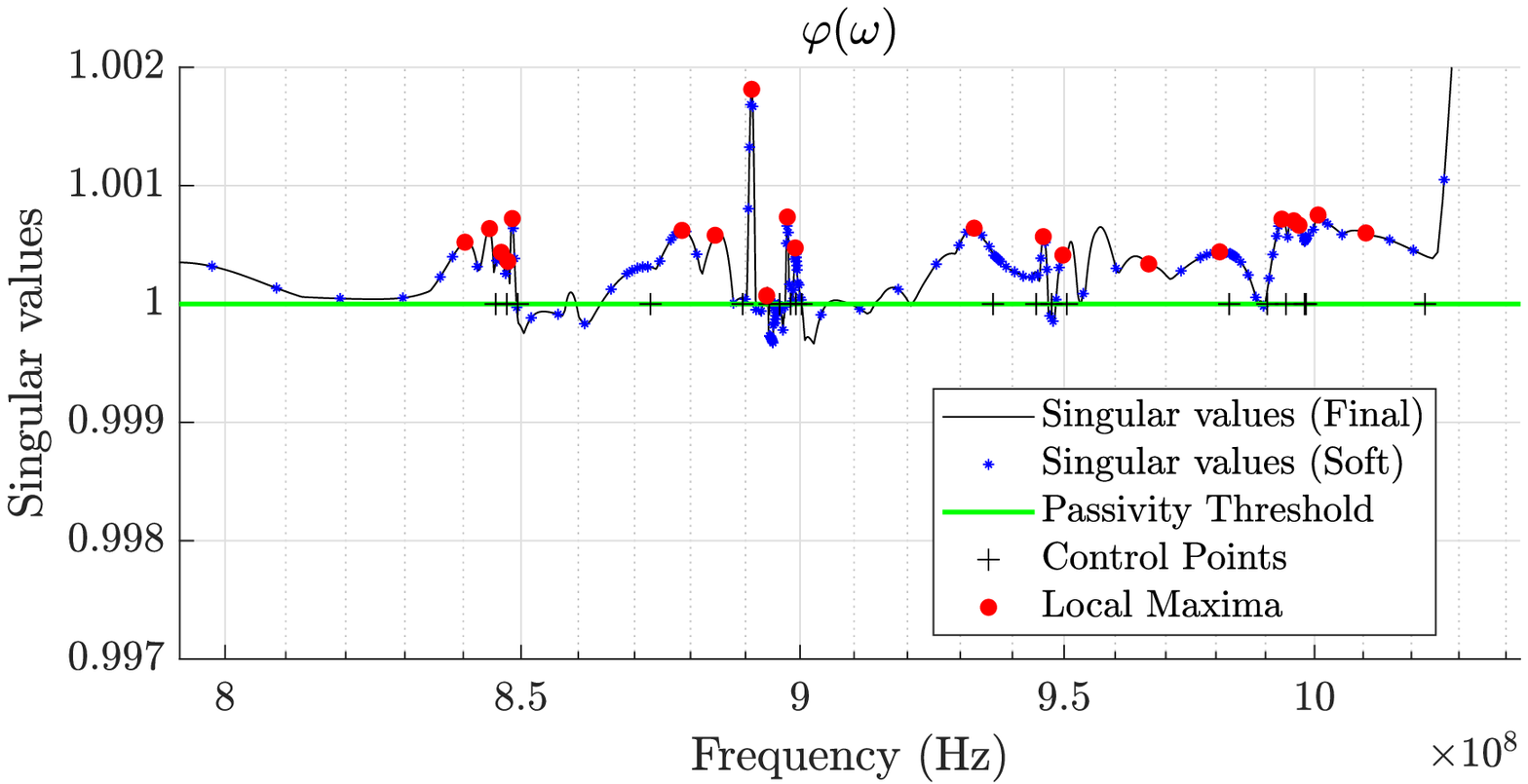}}
    \caption{Singular value plot of the shielding enclosure model. Bottom panel: enlarged view.}
    \label{fig:box_svals}
\end{figure}

We compare here the performance of proposed passivity characterization to various implementations of Hamiltonian-based tests available in the state-of-the-art commercial package IdEM~\cite{idem}. The structure that is selected for this comparison is a large-scale model of a shielding enclosure, depicted in Fig.~\ref{fig:box}. The enclosure is part of an energy-selective shield, obtained by loading the $P=20\times 20=400$ ports located throughout the opening with nonlinear elements (back-to-back diode pairs). For an overview of energy-selective shielding see~\cite{eleftheriades14_protec_weak_from_stron}, in particular~\cite{jnl-2020-temc-torben,Yang2013,yang2016impuls,yang2016design}.

The unloaded structure was initially characterized through a Method-of-Moments (MoM) frequency-domain code~\cite{concept}, obtaining a set of $P \times P$ scattering responses, each with a total of $500$ frequency samples over the frequency band $[0.25, 990]$~MHz. These samples were subjected to VF to obtain a rational model ($\bar{n}=26$ poles), leading to a state-space realization of size $N=10400$. Few selected port responses of the model are compared to field solver data in Figure~\ref{fig:box_responses}, demonstrating good model accuracy.

The model was processed by four different passivity characterization schemes, namely
\begin{itemize}
    \item Full Hamiltonian check: as implemented in~\cite{idem}, extracting all imaginary  eigenvalues of the full Hamiltonian matrix~\eqref{eq:Ham_scat_def};
    \item Sparse Hamiltonian check: as implemented in~\cite{idem}, based on the Hamiltonian eigensolver exploiting the iterative multishift Krylov subspace iterations of~\cite{jnl-2006-tadvp-LargePassivity};
    \item Adaptive Hamiltonian check: as implemented in~\cite{idem}, based on the hybrid sampling-sparse Hamiltonian eigensolver documented in~\cite{jnl-2007-tadvp-AdaptiveSampling};
    \item Proposed two-stage adaptive sampling check.
\end{itemize}
The Hamiltonian eigensolvers were available in two execution modes using $T=1$ and $T=8$ parallel threads, respectively. Proposed approach was implemented as a serial code, executed in a MATLAB~\cite{matlab} environment both enabling and disabling multithread capabilities (at low-level for SVD computations). The results are collected in Table~\ref{tab:boxIdEM}, which confirms a major speedup of proposed approach with respect to Hamiltonian characterizations. All methods provided consistent passivity characterization results in terms of number of violation bands. Figure~\ref{fig:box_svals} depicts the singular value trajectories computed by proposed adaptive sampling scheme, highlighting the local maxima exceeding the passivity threshold.

\begin{table}[t]
    \caption{Shielding enclosure: execution time in minutes for different passivity characterization approaches, under both single- and multi-threaded execution. See text for details.}
    \label{tab:boxIdEM}
    \centering
    \begin{tabular}{|c|c|c|c|c|c|c|}
    \hline
    & \multicolumn{3}{|c|}{Hamiltonian~\cite{idem}} & \multicolumn{3}{|c|}{Proposed} \\
    \hline
         \# Threads & Full & Sparse & Adaptive & \emph{soft} & \emph{hard} & \emph{final}\\
         \hline
        8 & 180 & 76 & 75 & 0.4 & 1.41 & 4.49\\
        \hline
        1 & 226 & 189 & 183 & 1.03 & 3.19 & 9.82\\
        \hline
    \end{tabular}
\end{table}

\subsection{A via array}\label{sec:via_array}

We analyze here a via array in a 8-layer Printed Circuit Board (PCB) structure. The array consists of $20 \times 20$ vias with a 4:1 signal to ground ratio, resulting in total of $P=640$ electrical ports (1--320 in the top layer, 321--640 in the bottom layer). A fast frequency-domain solver~\cite{6423302} was applied to extract the $P \times P$ scattering responses up to $30$~GHz, defining circular ports between via pad and antipad at top and bottom layers. These $150$ raw samples were in turn used to generate a large-scale macromodel through VF using the IdEM software~\cite{idem}, resulting in $\bar{n}=40$ pole-residue terms and a total number of states $N=25600$.

\begin{figure}
    \centerline{\includegraphics[width=\columnwidth]{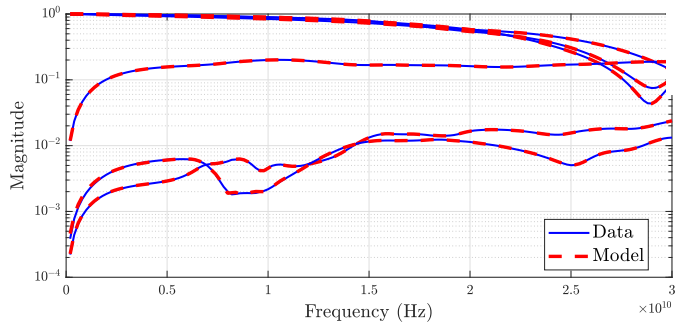}}
    \caption{Model vs data comparison of selected responses of the via array.}
    \label{fig:via_responses}
\end{figure}

\begin{figure}
    \centerline{\includegraphics[width=\columnwidth]{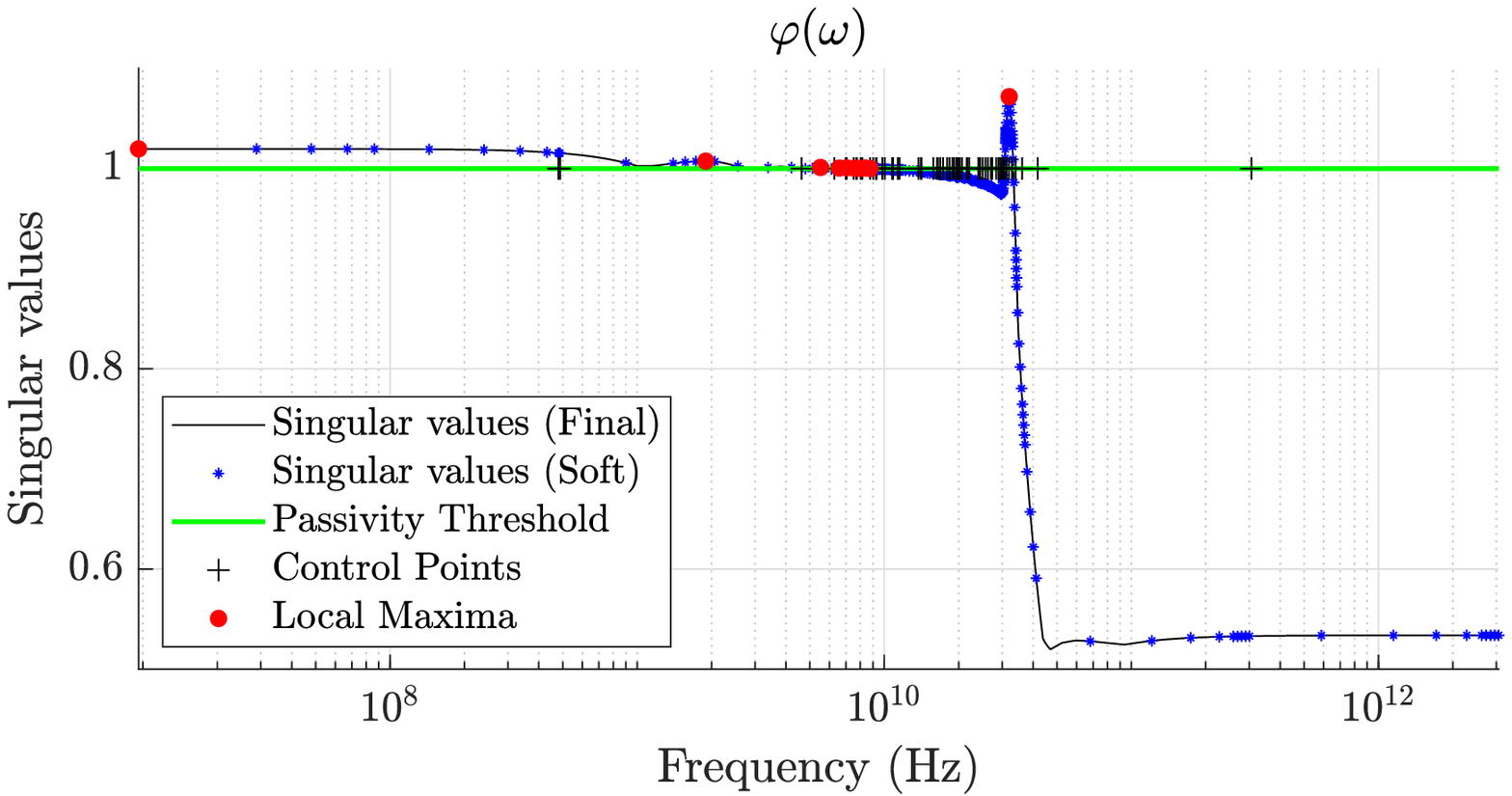}}
    \vspace*{4mm}
    \centerline{\includegraphics[width=\columnwidth]{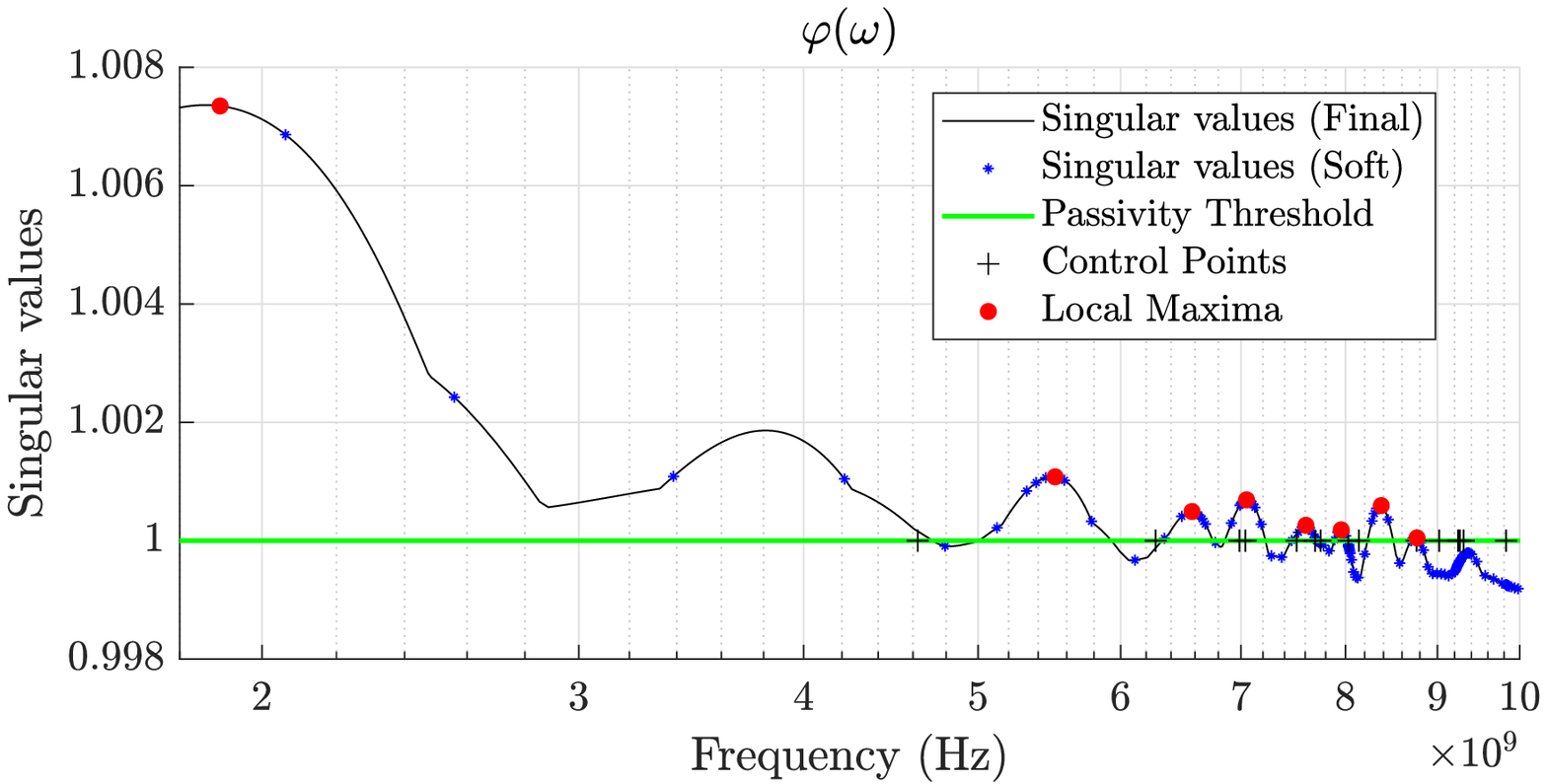}}
    \caption{Singular value plot of the via array model. Bottom panel: enlarged view showing that local maxima are correctly identified by proposed algorithm in \emph{hard} (not shown) and \emph{final} mode.}
    \label{fig:via_svals}
\end{figure}

The model was processed by the proposed passivity characterization algorithm, as well as to all Hamiltonian checks available in IdEM. Timing results are reported in Table~\ref{tab:via_array}, whereas Fig.~\ref{fig:via_responses} provides a comparison between selected model responses and corresponding field solver samples. Figure~\ref{fig:via_svals} depicts the largest singular value samples returned by proposed adaptive sampling scheme, highlighting all local maxima exceeding the passivity threshold. Execution in both \emph{hard} and \emph{final} mode identified all 11 local maxima, whereas the fast \emph{soft} mode found 10 maxima. Also for this example, the execution time provided by proposed passivity check algorithm is drastically reduced with respect to Hamiltonian checks, at least for \emph{soft} and \emph{hard} execution modes, with no loss of accuracy in the characterization of passivity violations. Full Hamiltonian checks could not be performed due to excessive storage requirements.

\begin{table}[t]
    \caption{Via array: execution time in minutes for different passivity characterization approaches, under both single- and multi-threaded execution. See text for details.}
    \label{tab:via_array}
    \centering
    \begin{tabular}{|c|c|c|c|c|c|c|}
    \hline
    & \multicolumn{3}{|c|}{Hamiltonian~\cite{idem}} & \multicolumn{3}{|c|}{Proposed} \\
    \hline
         \# Threads & Full & Sparse & Adaptive & \emph{soft} & \emph{hard} & \emph{final}\\
         \hline
        8 & Fail & 19 & 29 & 1.53 & 8.15 & 17.35 \\
        \hline
        1 & Fail & 67 & 68 & 4.41 & 23,1 & 47.05 \\
        \hline
    \end{tabular}
\end{table}

\subsection{Discussion}

The above numerical results show that the proposed algorithm appears to be the only viable and scalable solution for passivity characterization of large-sized models. The following considerations apply.
\begin{itemize}
    \item The number of subbands produced by the step-1 adaptive frequency warping grows with the number $\bar{n}$ of model poles and can be very large. This is not a limitation, since the independent processing of those subbands lends itself to a naive data-based parallelization, which is thus expected to provide even further speedup with respect to the already highly efficient implementation that is here documented. Code parallelization is left to future investigations.
    \item It may be the case that multiple passivity violation maxima in the same subband may be missed if the stopping thresholds are excessively loose or the allocated budget of evaluations is too small.
    \item In case a very accurate test is needed, asymptotic convergence to all local maxima can be guaranteed by releasing budget constraints and enabling also basket reuse for refinement, building on the provable asymptotic consistency of the NMSO algorithm class~\cite{2016_Dujaili_Suresh_NMSO}. We remark that this strategy was never required in our extensive tests, since proposed implementation never identified non-passive models as passive.
\end{itemize}

\section{Conclusion}\label{sec:conclusions}

This work introduced a two-stage hierarchical sampling-based passivity characterization algorithm for large-scale lumped macromodels.  The proposed scheme was tested on about 450 test cases of increasing complexity and reaching $P=640$ ports and more than $25\times 10^{3}$ states. The results show that the proposed approach outperforms Hamiltonian tests for medium and large-scale systems, with speedup in runtime exceeding $200\times$ in some cases.

Due to the multi-stage hierarchical implementation, our proposed algorithm was always in agreement with Hamiltonian checks. The only exception was a single case that was (correctly) detected as non-passive by proposed method, although the Hamiltonian check could not identify any passivity violation. Therefore, we conclude that proposed scheme is at least as reliable as state-of-the-art approaches, with much more favorable scalability with model size.

Significant research work is still needed for an efficient extraction and handling of large-scale macromodels. Fully coupled models with many ports and large dynamical orders are in fact not effective when synthesized as equivalent circuits and used in system-level simulations, due to the very large number of model parameters. Ad hoc model representations based on sparsification or compression appear to provide a better solution for speeding up numerical simulations, possibly combined with specialized solvers that are aware of the model structure. We are actively investigating in this direction, the results will be documented in a forthcoming report.

\section*{Acknowledgement}

The Authors are grateful to Morten Schierholz,  Institut f\"ur Theoretische Elektrotechnik,
Hamburg University of Technology, for providing the via array dataset of Section~\ref{sec:via_array}.

This work was supported in part by the German Research Foundation (DFG).

\bibliographystyle{IEEEtran}
\bibliography{IEEEabrv,PM_Biblio.bib,MyBib.bib}
	
\end{document}

%% file: PM_Macros.tex
\newcommand{\junk}[1] {}

\newcommand{\vet}[1]{\boldsymbol{#1}} 
\newcommand{\mat}[1]{\mathbf{#1}} 
\newcommand{\matg}[1]{\boldsymbol{#1}} 

\newcommand{\tran}{\mathsf{T}}

\newcommand{\eye}{\mathbb{I}}

\newcommand{\ham}[1]{{\matg{\cal #1}}}

\newcommand{\abs}[1]{\left\lvert #1 \right\rvert }

\def\XXint#1#2#3{{\setbox0=\hbox{$#1{#2#3}{\int}$}
\vcenter{\hbox{$#2#3$}}\kern-.5\wd0}}

\def\jj{\mathrm{j}}

\newcommand{\Real}{\mathbb{R}}
\newcommand{\Complex}{\mathbb{C}}



%% file: macrosFile.tex
\renewcommand{\vet}[1]{\boldsymbol{#1}} 
\renewcommand{\mat}[1]{\mathbf{#1}} 

\renewcommand{\tran}{\mathsf{T}}

\renewcommand{\eye}{\mat{I}}

\renewcommand{\abs}[1]{\left\lvert #1 \right\rvert }

\def\jj{\mathrm{j}}

\renewcommand{\Real}{\mathbb{R}}
\renewcommand{\Complex}{\mathbb{C}}


%% file: paramTable.tex
\begin{table}
	\caption{Control parameters of proposed step-1 frequency warping and normalization algorithm. Three different settings are suggested, to emphasize speed (\emph{soft}), accuracy (\emph{hard}) and to perform model qualification (\emph{final}).} 
	\label{table:parametersWarping}
		\centering
		\begin{tabular}{|c|c|c|c|c|c|c|c|c|}
			\hline
			\multicolumn{9}{|c|}{Step~1: Frequency Warping}\\
			\hline
			${\rm Mode}$ & $\rho$ & $R_{\rm cp}$ & $R_{\rm rp}$ & $R_{\rm hf}$ & $c$ & $Q_{\max}$ & $\kappa$ &  $d$ \\
			\hline
			\emph{soft} & $10^{3}$ & 1 & 2 & 5 & 50 & 500 & 3 & 0.5 \\
			\hline
			\emph{hard} & $\infty$ & 3 & 3 & 6 & 50 & 500 & 3 & 0.5 \\
			\hline
			\emph{final} & $\infty$ & 3 & 3 & 6 & 50 & 500 & 3 & 0.5 \\
			\hline
		\end{tabular}
\end{table}
\begin{table}[t]
	\caption{Control parameters of proposed step-2 adaptive sampling algorithm. Three different settings are suggested, to emphasize speed (\emph{soft}), accuracy (\emph{hard}) and to perform model qualification (\emph{final}).} 
	\label{table:parametersNMSO}
		\centering
			\begin{tabular}{|c|c|c|c|c|c|c|}
				\hline
				\multicolumn{7}{|c|}{Step~2: Modified NMSO}\\
				\hline
				${\rm Mode}$ &$M$ & $\delta\omw$ , $\delta\ssr$ &  $\delta\eta$ & $\varepsilon$ & $\varrho$& $\budget$ \\
				\hline
				\emph{soft} &5 & $10^{-8}$ & $10^{-3}$ & $10^{-3}$ & 0.1 & $7,10,20,\dots,100$ \\
				\hline
				\emph{hard} & 5 & $10^{-8}$ & $10^{-2}$ & $10^{-3}$ & 0.1 & $10,20,\dots,100$\\
				\hline
				\emph{final} &3 & $10^{-8}$ & $10^{-3}$ & $10^{-4}$ & 0.1 & $50, 100,\dots,250$\\
				\hline
			\end{tabular}
\end{table}

%% file: main_passCheck.bbl
\begin{thebibliography}{10}
\providecommand{\url}[1]{#1}
\csname url@samestyle\endcsname
\providecommand{\newblock}{\relax}
\providecommand{\bibinfo}[2]{#2}
\providecommand{\BIBentrySTDinterwordspacing}{\spaceskip=0pt\relax}
\providecommand{\BIBentryALTinterwordstretchfactor}{4}
\providecommand{\BIBentryALTinterwordspacing}{\spaceskip=\fontdimen2\font plus
\BIBentryALTinterwordstretchfactor\fontdimen3\font minus
  \fontdimen4\font\relax}
\providecommand{\BIBforeignlanguage}[2]{{%
\expandafter\ifx\csname l@#1\endcsname\relax
\typeout{** WARNING: IEEEtran.bst: No hyphenation pattern has been}%
\typeout{** loaded for the language `#1'. Using the pattern for}%
\typeout{** the default language instead.}%
\else
\language=\csname l@#1\endcsname
\fi
#2}}
\providecommand{\BIBdecl}{\relax}
\BIBdecl

\bibitem{tpwrd-1999-VF}
B.~Gustavsen and A.~Semlyen, ``Rational approximation of frequency domain
  responses by vector fitting,'' \emph{Power Delivery, IEEE Transactions on},
  vol.~14, no.~3, pp. 1052--1061, jul 1999.

\bibitem{mon-2015-PM}
S.~Grivet-Talocia and B.~Gustavsen, \emph{Passive Macromodeling: Theory and
  Applications}.\hskip 1em plus 0.5em minus 0.4em\relax New York: John Wiley
  and Sons, 2016 (published online on Dec 7, 2015).

\bibitem{jnl-2007-tadvp-Fundamentals}
P.~Triverio, S.~Grivet-Talocia, M.~S. Nakhla, F.~Canavero, and R.~Achar,
  ``Stability, causality, and passivity in electrical interconnect models,''
  \emph{{IEEE} Trans. Advanced Packaging}, vol.~30, no.~4, pp. 795--808, Nov
  2007.

\bibitem{Willems_dissipative}
\BIBentryALTinterwordspacing
J.~C. Willems, ``\BIBforeignlanguage{English}{Dissipative dynamical systems
  part {I}: General theory},'' \emph{\BIBforeignlanguage{English}{Archive for
  Rational Mechanics and Analysis}}, vol.~45, no.~5, pp. 321--351, 1972.
  [Online]. Available: \url{http://dx.doi.org/10.1007/BF00276493}
\BIBentrySTDinterwordspacing

\bibitem{Wohlers}
M.~R. Wohlers, \emph{Lumped and Distributed Passive Networks}.\hskip 1em plus
  0.5em minus 0.4em\relax Academic press, 1969.

\bibitem{Anderson}
B.~D.~O. Anderson and S.~Vongpanitlerd, \emph{Network analysis and
  synthesis}.\hskip 1em plus 0.5em minus 0.4em\relax Prentice-Hall, 1973.

\bibitem{1262465}
C.~P. Coelho, J.~Phillips, and L.~M. Silveira, ``A convex programming approach
  for generating guaranteed passive approximations to tabulated
  frequency-data,'' \emph{Computer-Aided Design of Integrated Circuits and
  Systems, IEEE Transactions on}, vol.~23, no.~2, pp. 293 -- 301, feb. 2004.

\bibitem{4114368}
D.~Saraswat, R.~Achar, and M.~S. Nakhla, ``Fast passivity verification and
  enforcement via reciprocal systems for interconnects with large order
  macromodels,'' \emph{Very Large Scale Integration (VLSI) Systems, IEEE
  Transactions on}, vol.~15, no.~1, pp. 48--59, Jan 2007.

\bibitem{1498835}
------, ``Global passivity enforcement algorithm for macromodels of
  interconnect subnetworks characterized by tabulated data,'' \emph{Very Large
  Scale Integration (VLSI) Systems, IEEE Transactions on}, vol.~13, no.~7, pp.
  819--832, July 2005.

\bibitem{5715905}
C.~S. Saunders, J.~Hu, C.~E. Christoffersen, and M.~B. Steer, ``Inverse
  singular value method for enforcing passivity in reduced-order models of
  distributed structures for transient and steady-state simulation,''
  \emph{Microwave Theory and Techniques, IEEE Transactions on}, vol.~59, no.~4,
  pp. 837--847, April 2011.

\bibitem{6298986}
T.~Brull and C.~Schroder, ``Dissipativity enforcement via perturbation of
  para-{H}ermitian pencils,'' \emph{Circuits and Systems I: Regular Papers,
  IEEE Transactions on}, vol.~60, no.~1, pp. 164--177, Jan 2013.

\bibitem{5484629}
S.~Gao, Y.-S. Li, and M.-S. Zhang, ``An efficient algebraic method for the
  passivity enforcement of macromodels,'' \emph{Microwave Theory and
  Techniques, IEEE Transactions on}, vol.~58, no.~7, pp. 1830--1839, July 2010.

\bibitem{4492758}
T.~D'haene and R.~Pintelon, ``Passivity enforcement of transfer functions,''
  \emph{Instrumentation and Measurement, IEEE Transactions on}, vol.~57,
  no.~10, pp. 2181--2187, Oct 2008.

\bibitem{4526214}
B.~Porkar, M.~Vakilian, R.~Iravani, and S.~M. Shahrtash, ``Passivity
  enforcement using an infeasible-interior-point primal-dual method,''
  \emph{Power Systems, IEEE Transactions on}, vol.~23, no.~3, pp. 966--974, Aug
  2008.

\bibitem{6224200}
T.~Wang and Z.~Ye, ``Robust passive macro-model generation with local
  compensation,'' \emph{Microwave Theory and Techniques, IEEE Transactions on},
  vol.~60, no.~8, pp. 2313--2328, Aug 2012.

\bibitem{jnl-2004-cas-Hamiltonian}
S.~Grivet-Talocia, ``Passivity enforcement via perturbation of {H}amitonian
  matrices,'' \emph{{IEEE} Trans. Circuits and Systems I: Fundamental Theory
  and Applications}, vol.~51, no.~9, pp. 1755--1769, September 2004.

\bibitem{jnl-2006-tadvp-LargePassivity}
S.~Grivet-Talocia and A.~Ubolli, ``On the generation of large passive
  macromodels for complex interconnect structures,'' \emph{{IEEE} Trans.
  Advanced Packaging}, vol.~29, no.~1, pp. 39--54, February 2006.

\bibitem{jnl-2007-tadvp-AdaptiveSampling}
S.~Grivet-Talocia, ``An adaptive sampling technique for passivity
  characterization and enforcement of large interconnect macromodels,''
  \emph{{IEEE} Trans. Advanced Packaging}, vol.~30, no.~2, pp. 226--237, May
  2007.

\bibitem{jnl-2008-tadvp-PassivityMethods}
S.~Grivet-Talocia and A.~Ubolli, ``A comparative study of passivity enforcement
  schemes for linear lumped macromodels,'' \emph{{IEEE} Trans. Advanced
  Packaging}, vol.~31, no.~4, pp. 673--683, Nov 2008.

\bibitem{jnl-2009-ijcta-destabilize}
S.~Grivet-Talocia, ``On driving non-passive macromodels to instability,''
  \emph{International Journal of Circuit Theory and Applications}, vol.~37,
  no.~8, pp. 863--886, Oct 2009.

\bibitem{hir2019}
\BIBentryALTinterwordspacing
``{Heterogeneous Integration Roadmap, 2019 Edition}.'' [Online]. Available:
  \url{https://eps.ieee.org/technology/heterogeneous-integration-roadmap/2019-edition.html}
\BIBentrySTDinterwordspacing

\bibitem{eleftheriades14_protec_weak_from_stron}
\BIBentryALTinterwordspacing
G.~V. Eleftheriades, ``Protecting the weak from the strong,'' \emph{Nature},
  vol. 505, no. 7484, pp. 490--491, 2014. [Online]. Available:
  \url{https://doi.org/10.1038/nature12852}
\BIBentrySTDinterwordspacing

\bibitem{jnl-2020-temc-torben}
T.~{Wendt}, C.~{Yang}, H.~{D. Br{\"u}ns}, S.~{Grivet-Talocia}, and
  C.~{Schuster}, ``A macromodeling-based hybrid method for the computation of
  transient electromagnetic fields scattered by nonlinearly loaded metal
  structures,'' \emph{IEEE Transactions on Electromagnetic Compatibility},
  vol.~62, no.~4, pp. 1098--1110, Aug 2020.

\bibitem{Yang2013}
C.~Yang, P.~Liu, and X.~Huang, ``{A novel method of energy selective surface
  for adaptive HPM/EMP protection},'' \emph{IEEE Antennas Wirel. Propag.
  Lett.}, vol.~12, pp. 112--115, 2013.

\bibitem{wendt19_num_comp_study}
T.~Wendt, C.~Yang, S.~Grivet-Talocia, and C.~Schuster, ``Numerical complexity
  study of solving hybrid multiport field-circuit problems for diode grids,''
  in \emph{International Conference on Electromagnetics in Advanced
  Applications (ICEAA 2019)}, September 2019.

\bibitem{yang2016impuls}
C.~Yang, H.-D. Br{\"u}ns, P.~Liu, and C.~Schuster, ``Impulse response
  optimization of band-limited frequency data for hybrid field-circuit
  simulation of large-scale energy-selective diode grids,'' \emph{IEEE Trans.
  Electromagn. Compat.}, vol.~PP, no.~99, pp. 1--9, May 2016.

\bibitem{yang2016design}
C.~Yang, P.~Liu, H.-D. Br{\"u}ns, and C.~Schuster, ``Design aspects for {HIRF}
  protection of a rectangular metallic cavity using energy selective diode
  grids,'' in \emph{2016 Asia-Pacific International Symposium on
  Electromagnetic Compatibility (APEMC)}, vol.~1.\hskip 1em plus 0.5em minus
  0.4em\relax IEEE, 2016, pp. 35--37.

\bibitem{mwcl-2008-Deschrijver-FastVF}
D.~Deschrijver, M.~Mrozowski, T.~Dhaene, and D.~{De Zutter}, ``Macromodeling of
  multiport systems using a fast implementation of the vector fitting method,''
  \emph{Microwave and Wireless Components Letters, IEEE}, vol.~18, no.~6, pp.
  383--385, june 2008.

\bibitem{jnl-2012-tcpmt-comp}
S.~B. Olivadese and S.~Grivet-Talocia, ``Compressed passive macromodeling,''
  \emph{IEEE Transactions on Components, Packaging and Manufacturing
  Technology}, vol.~2, no.~8, pp. 1378--1388, August 2012.

\bibitem{jnl-2011-tcpmt-pvf}
A.~Chinea and S.~Grivet-Talocia, ``On the parallelization of vector fitting
  algorithms,'' \emph{IEEE Transactions on Components, Packaging and
  Manufacturing Technology}, vol.~1, no.~11, pp. 1761--1773, November 2011.

\bibitem{9123948}
S.~{Ganeshan}, N.~K. {Elumalai}, R.~{Achar}, and W.~K. {Lee}, ``Gvf: Gpu-based
  vector fitting for modeling of multiport tabulated data networks,''
  \emph{IEEE Transactions on Components, Packaging and Manufacturing
  Technology}, vol.~10, no.~8, pp. 1375--1387, 2020.

\bibitem{jnl-2013-tcpmt-parallel}
\BIBentryALTinterwordspacing
A.~Chinea, S.~Grivet-Talocia, S.~B. Olivadese, and L.~Gobbato,
  ``High-performance passive macromodeling algorithms for parallel computing
  platforms,'' \emph{IEEE Transactions on Components, Packaging, and
  Manufacturing Technology}, vol.~3, no.~7, pp. 1188--1203, July 2013.
  [Online]. Available:
  \url{http://ieeexplore.ieee.org/stamp/stamp.jsp?tp=&arnumber=6510533}
\BIBentrySTDinterwordspacing

\bibitem{BBK89}
S.~Boyd, V.~Balakrishnan, and P.~Kabamba, ``A bisection method for computing
  the {$H_\infty$} norm of a transfer matrix and related problems,''
  \emph{Mathematics of Control, Signals and Systems}, vol.~2, no.~3, pp.
  207--219, 1989.

\bibitem{2016_Dujaili_Suresh_NMSO}
\BIBentryALTinterwordspacing
A.~Al-Dujaili and S.~Suresh, ``{A Naive multi-scale search algorithm for global
  optimization problems},'' \emph{Information Sciences}, vol. 372, pp.
  294--312, dec 2016. [Online]. Available:
  \url{https://linkinghub.elsevier.com/retrieve/pii/S0020025516305370}
\BIBentrySTDinterwordspacing

\bibitem{1046889}
B.~Gustavsen, ``Computer code for rational approximation of frequency dependent
  admittance matrices,'' \emph{Power Delivery, IEEE Transactions on}, vol.~17,
  no.~4, pp. 1093--1098, oct 2002.

\bibitem{gilbert1963}
E.~G. Gilbert, ``Controllability and observability in multivariable control
  systems,'' \emph{Journal of the Society for Industrial \& Applied
  Mathematics, Series A: Control}, vol.~1, no.~2, pp. 128--151, 1963.

\bibitem{scherer2000linear}
C.~Scherer and S.~Weiland, ``Linear matrix inequalities in control,''
  \emph{Lecture Notes, Dutch Institute for Systems and Control, Delft, The
  Netherlands}, 2000.

\bibitem{KYP-IPM}
L.~Vandenberghe, V.~Balakrishnan, R.~Wallin, A.~Hansson, and T.~Roh,
  \emph{Interior-Point Algorithms for Semidefinite Programming Problems Derived
  from the KYP Lemma}.\hskip 1em plus 0.5em minus 0.4em\relax Springer, Berlin,
  Heidelberg, 05 2005, vol. 312, pp. 579--579.

\bibitem{1250052}
H.~Chen and J.~Fang, ``Enforcing bounded realness of {S} parameter through
  trace parameterization,'' in \emph{Electrical Performance of Electronic
  Packaging, 2003}, Oct 2003, pp. 291--294.

\bibitem{5456981}
Z.~Ye, L.~M. Silveira, and J.~R. Phillips, ``Extended {H}amiltonian pencil for
  passivity assessment and enforcement for {S}-parameter systems,'' in
  \emph{Design, Automation Test in Europe Conference Exhibition (DATE), 2010},
  March 2010, pp. 1148--1152.

\bibitem{idem}
\BIBentryALTinterwordspacing
``{IdEM R2018}, {{Dassault Syst\`emes}}.'' [Online]. Available:
  \url{www.3ds.com/products-services/simulia/products/idem/}
\BIBentrySTDinterwordspacing

\bibitem{concept}
\BIBentryALTinterwordspacing
``{CONCEPT-II, 2019}.'' [Online]. Available:
  \url{http://www.tet.tuhh.de/concept/}
\BIBentrySTDinterwordspacing

\bibitem{matlab}
\emph{{MATLAB} User's Guide}, The Mathworks, Inc., {\tt www.mathworks.com}.

\bibitem{6423302}
S.~{M\"uller}, F.~{Happ}, X.~{Duan}, R.~{Rimolo-Donadio}, H.~{Br\"uns}, and
  C.~{Schuster}, ``Complete modeling of large via constellations in multilayer
  printed circuit boards,'' \emph{IEEE Transactions on Components, Packaging
  and Manufacturing Technology}, vol.~3, no.~3, pp. 489--499, 2013.

\end{thebibliography}
